\documentclass[twocolumn]{aastex63}

\usepackage{graphicx}
\usepackage{amsmath}
\usepackage{amssymb}
\usepackage{natbib}
\usepackage{color}
\usepackage{enumitem}
\usepackage[T1]{fontenc}

\usepackage{txfonts}
\begin{document}

\submitjournal{AJ}

\shorttitle{Gas and dust dynamics}
\shortauthors{Flock et al.}

\title{Gas and dust dynamics in starlight-heated protoplanetary disks}

\correspondingauthor{Mario Flock}
\email{flock@mpia.de}

\author[0000-0002-9298-3029]{Mario Flock}
\affiliation{Max-Planck Institute for Astronomy (MPIA), K\"{o}nigstuhl 17, 69117 Heidelberg, Germany}

\author[0000-0001-8292-1943]{Neal J. Turner}
\affiliation{Jet Propulsion Laboratory, California Institute of Technology, Pasadena, California 91109, USA}

\author{Richard P. Nelson}
\affiliation{Astronomy Unit, Queen Mary University of London, Mile End Road, London E1 4NS, UK}

\author{Wladimir Lyra}
\affiliation{Department of Astronomy, New Mexico State University, MSC 4500, Las Cruces, NM 88003, USA}

\author{Natascha Manger}
\affiliation{Center for Computational Astrophysics, Flatiron Institute, 162
  Fifth Ave, New York, NY 10010, USA}
  
\author{Hubert Klahr}
\affiliation{Max-Planck Institute for Astronomy (MPIA), K\"{o}nigstuhl 17, 69117 Heidelberg, Germany}

\begin{abstract}Theoretical models of the ionization state in protoplanetary
  disks suggest the existence of large areas with low ionization and weak
  coupling between the gas and magnetic fields. In this regime hydrodynamical
  instabilities may become important. In this work we investigate the gas and
  dust structure and dynamics for a typical T Tauri system under the influence
  of the vertical shear instability (VSI). We use global 3D radiation
  hydrodynamics simulations covering all $360^\circ$ of azimuth with embedded
  particles of 0.1 and 1mm size, evolved for 400 orbits. Stellar irradiation
  heating is included with opacities for 0.1- to 10-$\mu$m-sized
  dust. Saturated VSI turbulence produces a stress-to-pressure ratio of
  $\alpha \simeq 10^{-4}$. The value of $\alpha$ is lowest within 30~au of the
  star, where thermal relaxation is slower relative to the orbital period and
  approaches the rate below which VSI is cut off.  The rise in $\alpha$ from
  20 to 30~au causes a dip in the surface density near 35~au, leading to
  Rossby wave instability and the generation of a stationary, long-lived
  vortex spanning about 4~au in radius and 40~au in azimuth. Our results
  confirm previous findings that mm size grains are strongly vertically mixed
  by the VSI. The scale height aspect ratio for 1mm grains is determined to be
  0.037, much higher than the value $H/r=0.007$ obtained from millimeter-wave
  observations of the HL~Tau system. The measured aspect ratio is better fit
  by non-ideal MHD models. In our VSI turbulence model, the mm grains drift radially inwards and many are trapped and concentrated inside the
  vortex. The turbulence induces a velocity dispersion of $\sim 12$~m/s for
  the mm grains, indicating that grain-grain collisions could lead to
  fragmentation.
\end{abstract}


\section{Introduction}
The dynamical states of protostellar disks during the first million years
after their formation are crucial for understanding planet
formation processes. As the coupling of the magnetic field to the gas in these
dense and cold disks is believed to be weak \citep{tur14,arm19}, it seems likely that various
hydrodynamical instabilities play important roles,
such as the Goldreich$-$Schubert$-$Fricke (GSF) instability \citep{gol67,fri68}, the convective instability \citep{cam73}, the Papaloizou$-$Pringle instability \citep{pap84}, the baroclinic instability \citep{kla03,les10}, 
the convective overstability \citep{lyr14,kla14}, the zombie vortex instability \citep{mar13,mar15,les16,umu16b} or the spiral wave instability \citep{bae16}. %
In recent years, it has become clear that hydrodynamical instabilities might play important roles
in explaining the structure and evolution of protoplanetary disks
\citep{nel13,lin15,sto16,flo17,man18,har18}, while the more general role played by hydrodynamical instabilities in driving
angular momentum transport remains unresolved \citep{fro19}. A special focus
has been placed on the GSF instability, later renamed for disks to the vertical shear
instability (VSI), which is believed
to operate in regions where the magnetic coupling is low and the thermal
equilibrium timescale is much shorter than the dynamical timescale
\citep{urp98,nel13,lin15,ric16,mal17,lat18,pfe19,lyr19}.
Recent hydrodynamical simulations have found that the VSI develops in the
non-linear regime to produce vortices \citep{ric16,man18} due to 
the Rossby-Wave-Instability \citep{li00}. Also, works on
the VSI have found stress-to-pressure ratios $\alpha$ (measuring the $r - \phi$ component of
the Reynolds stress tensor) ranging from $\alpha=10^{-6}$ to $\alpha=10^{-3}$ \citep{nel13,sto14,ric16,flo17,man18}. Of special interest for planet formation models are the dust dynamics, and the spatial
structure of the dust distribution, which are driven by VSI turbulence. Continuum observations by the Atacama
Large Millimeter Array (ALMA) which trace the dust thermal emission have revealed
astonishingly thin dust structures in the vertical dimension \citep{par15,pin16,vanb17,liu17}. One way to
explain these thin dust components is a low level of disk turbulence. In our recent work
we showed that the magneto-rotational instability operating in the upper
layers of the disk, providing only mild perturbations to the flow in the midplane, would be able to explain the required degree of vertical mixing \citep{flo15,flo17}. In contrast, we showed that the characteristic vertical motions
of the vertical shear instability efficiently lift up mm size grains
\citep{flo17}, leading to a vertical dust distribution that is much more
spatially extended than inferred from the observations. Previous work also
found that the strength of this vertical mixing is much greater than expected from the stress-to-pressure
ratio $\alpha$, because of the inherently anisotropic nature of the turbulence
generated by the VSI \citep{nel13,sto17}.

Several previous works have demonstrated the appearance of vortices driven by hydrodynamical
instabilities. Examples include the convective overstability \citep{lyr14,kla14}
and the non-linear subcritical baroclinic instability (SBI)
\citep{les10}. Both of these instabilities require a thermal relaxation time
close to the inverse orbital frequency $\Omega^{-1}$. For
shorter thermal relaxation timescales, as present in our model, the SBI does not operate
\citep{les10,bar16}, while the growth of the convective overstability is
very slow \citep{kla14}. In the context of the vertical shear instability,
\citet{ric16} described the appearance of small scale vortices in simulations of the VSI, driven by radially-narrow vorticity perturbations that underwent local Rossby Wave Instability. In models with very short thermal relaxation time scales, \citet{ric16} found that the 
aspect ratios of the vortices was small (typically $\chi < 2$) because of the
strength of the VSI-induced vorticity perturbations, and they survived for
only a few orbits at most. \citet{man18} reported also the appearance of several vortices in their 3D isothermal
hydrodynamical simulations, however of much larger scales and most likely
caused by the Rossby Wave instability.

In this work we want to re-investigate the turbulent properties of the
vertical shear instability using high-resolution radiation
hydrodynamical simulations covering all $360^\circ$ of azimuth with stellar irradiation, and the consequences
of the turbulence for the evolution and distribution of embedded dust grains.
Our radiation hydrodynamic stratified disk simulations treat irradiation by
the central star, include realistic dust opacities and achieve a resolution of more than 70 cells
per disk scale height. The dynamics and structure of the dust are studied with
one million Lagrangian dust particles embedded in the disk. %\citet{bar15} pointed out the importance of a large vertical extent and high resolution to capture the fastest growing modes at the disc surfaces. 
The paper is laid out as follows. In Section~2 we present the
numerical method and the disk model, and in Section~3 we present the simulation
results for the gas and dust. In Section~4 we present a discussion of the results and
in Section~5 we draw our conclusions.

\section{Numerical method and disk model}
The radiation hydrodynamic equations are solved using the hybrid stellar irradiation and flux-limited diffusion method developed by \citet{flo13} as implemented in
version 4.2 of the PLUTO code \citep{mig07}. The numerical setup
follows \citep{flo17} in using the piece-wise parabolic method (PPM), the Harten-Lax-Van Leer approximate Riemann solver with the contact discontinuity (HLLC), the FARGO scheme \citep{mas00,mig12a} and the Runge-Kutta time integrator. The Courant number is set to 0.3. The equations solved are: 
\begin{eqnarray}
 \frac{\partial \mathrm{\rho}}{\partial \mathrm{t} } + \nabla \cdot \left [  \mathrm{\rho} \vec{v}\right ] &=&
0 \, , \label{eq:MDH_RHO} \\
 \frac{\partial \mathrm{\rho} \vec{v}}{\partial \mathrm{t}} + \nabla \cdot \left [
  \mathrm{\rho} \vec{v} \vec{v}^T \right ] + \rm \nabla \mathrm{P}
&=& - \mathrm{\rho} \nabla \mathrm{\Phi} \, , \label{eq:MDH_MOM} \\
 \frac{\partial \mathrm{E}}{\partial \mathrm{t}} + \nabla \cdot \left [ (\mathrm{E} + \mathrm{P})\vec{v} \right ]  &=& - \mathrm{\rho} \vec{v} \cdot \nabla \mathrm{\Phi} \nonumber \\& &  - \kappa_\mathrm{P}(\mathrm{T}) \mathrm{\rho} \mathrm{c} (\rm \mathrm{a_R} \mathrm{T}^4 - \mathrm{E_R} )\nonumber\\ & & - \nabla \cdot \mathrm{F}_* \, , \label{eq:MDH_EN} \\
\frac{\partial \mathrm{E_R}}{\partial \mathrm{t}} - \nabla \frac{ \mathrm{c} \mathrm{\lambda}}{ \kappa_\mathrm{R}(\mathrm{T}) \rho} \nabla \mathrm{E_R} &=& \kappa_\mathrm{P}(\mathrm{T}) \mathrm{\rho} \mathrm{c} ( \mathrm{a_R} \mathrm{T}^4 - \mathrm{E_R}) \, , \label{eq:ER} 
\end{eqnarray}
with the density $\mathrm{\rho}$, velocity vector $\vec{v}$, gas pressure 
\begin{equation}
\mathrm{P}= \frac{\mathrm{\rho} \mathrm{k_B} \mathrm{T}} { \mathrm{\mu_g} \mathrm{u}},
\end{equation}
gas temperature $\mathrm{T}$, mean molecular weight $\mathrm{\mu_g}$,
Boltzmann constant $\mathrm{k_B}$, atomic mass unit $\mathrm{u}$, total energy
per unit volume $\mathrm{E}= \mathrm{\rho} \mathrm{\epsilon} + 0.5
\mathrm{\rho} \vec{v}^{\,2}$  and gas internal energy per unit volsume
$\mathrm{\rho} \mathrm{\epsilon}$. The closure relation between gas pressure
and internal energy is provided by $\mathrm{P}=(\mathrm{\Gamma} -1)
\mathrm{\rho} \mathrm{\epsilon}$ with the adiabatic index
$\mathrm{\Gamma}$. Other symbols include the radiation energy E$_\mathrm{R}$,
the stellar irradiation flux $\mathrm{F}_*$, the Rosseland and
Planck mean opacities $\mathrm{\kappa_R}$ and $\mathrm{\kappa_P}$, the
radiation constant $\mathrm{a_R}=4 \mathrm{\sigma_b}/\mathrm{c}$, the
Stefan-Boltzmann constant $\mathrm{\sigma_b}$, and the speed of light $c$. The flux limiter 
\begin{equation}
\mathrm{\lambda} = \frac{2 + \mathrm{R}}{6 + 3\mathrm{R} + \mathrm{R}^2},
\end{equation}
is taken from \citet[][ Eq.~28 therein]{lev81} with the inverse optical depth
per radiation energy scale length 
\begin{equation}
\mathrm{R} = \frac{|\nabla \mathrm{E_R} |}{\mathrm{\kappa_R \rho} \mathrm{E_R}}.
\end{equation}
The gas is a mixture of molecular hydrogen, helium, and heavier elements with solar abundances \citep{dec78} so that $\mathrm{\mu_g} =2.35$ and $\mathrm{\Gamma}=1.42$. In this work we consider the frequency integrated irradiation flux at a radial distance $\mathrm{r}$ to be
\begin{equation}
\rm F_*(r) = \left (  \frac{R_*}{r}\right )^2  \sigma_b T_*^4 e^{-\tau}, 
\label{eq:IRRAD}
\end{equation}
with $\rm T_*$ and $\rm R_*$ being the surface temperature
and radius of the star. The radial optical depth to the irradiation flux at
meridional angle $\theta$ is given by:
\begin{equation}
\rm \tau(r,\theta)=\int_{R_*}^r \kappa(T_*) \rho_{dust}(r,\theta)  dr = \tau_0 + \int_{R_0}^r\kappa(T_*) \rho_{dust}(r,\theta)  dr \, ,
\label{eq:TAU}
\end{equation}
where $\rm R_0$ denotes the inner radius of the computational domain. The
quantity $\rm \tau_0$ is the inner optical depth provided by material located
between the position where the stellar magnetic field truncates the disk at
$\rm 3 R_*$ and the innermost radial position of our simulation domain $\rm
R_0$. We choose $\rm \tau_{0}=\kappa_* \rho_{R_0} (R_0-3R_*)$.

As in the previous work we use two opacities, one to the starlight
$\kappa_*=1300$~cm$^2$~g$^{-1}$, and the other to the disk's thermal
re-emission $\kappa_d=400$~cm$^2$~g$^{-1}$. These are Planck-weighted averages
of the frequency-dependent dust opacity, at respectively the stellar
temperature and a typical re-emission color temperature of 300~K.
The starlight opacity exceeding the thermal re-emission opacity yields a warm
surface layer that is opaque to the starlight but optically-thin to its own
emission \citep{chi97}. To be able to compare the results with our previous work we use
the same opacities as in \citet{flo17} and set $\kappa_R=\kappa_P$.
The frequency-dependent opacity is calculated using Mie theory for spheres of
silicate and graphite with a power-law radius distribution of exponent -3.5
between 0.1 and 10~$\mu$m \citep{flo17}. 

For all models we use a modified outflow boundary condition, preventing inflow
at the radial and $\theta$ boundaries. In addition, we extrapolate the
density logarithmically into the meridional ghost zones. In the radial
direction we apply buffer zones, relaxing the surface density to the initial
profile in the region 20-22~au and damping the radial velocity to zero at
20-21 and 97-100~au. Without these buffers, mass loss would lower and reshape
the starlight-absorbing layer, altering the temperature distribution. The buffer zones are excluded when we present our kinematic analysis of the results.

\subsection{Initial condition and run procedure}
\label{sec:IC}
The initial condition is derived by constructing 2D axisymmetric profiles of
density, temperature and rotation velocity which are in radiation hydrostatic
equilibrium. We use star and disk parameters from a model of a typical T
Tauri system which has been applied to explain various spatially-resolved multi-band observations of protostellar disks \citep{wol03,wol08,sch08,sch09,sau09,mad12,liu12,gra13}.
In Table~\ref{tab:model} we summarize the parameters that define the setup. As presented in \citet{flo16}, we construct the initial condition by iterating between the
radiation transfer and solving for hydrostatic
equilibrium. The opacity is determined the by the dust-to-gas mass ratio of
sub-10-$\mu$m grains which we take to be $10^{-3}$, the same as in model
\texttt{M3} of \citet{flo17}. This choice yields thermal equilibration
timescales short enough for the VSI to operate as detailed below in
sec. 4.2. To obtain the model \texttt{M3}, \citet{flo17} extended the 2D axisymmetric
initial conditions to 3D with an azimuthal extent of $\pi/8$ radians. Small velocity perturbations were added and the simulation
was run for 400 inner orbits. 

To save recomputing the linear growth of the VSI, we here extend and restart the reference model
\texttt{M3} after 400 inner orbits. We first stretch the azimuthal extent by a
factor 2 and then repeat it 8 times to fill the new domain's $2\pi$-radian
azimuthal extent.  The stretching keeps the number of azimuthal cells low
enough that the grid fits in the available memory. We next add small cellwise
random velocities with amplitude $10^{-4}c_s$ to the
existing velocity field to break the eight-fold symmetry.

The logarithmically-spaced radial and uniformly-spaced meridional and
azimuthal grids yield cells with aspect ratio approximately 1:1:2 and a resolution of
around 70 cells per scale height. As model \texttt{M3} produced
features that were mostly axisymmetric, reducing the azimuthal resolution a
factor two is unlikely to influence the results. The simulation with the new domain is then evolved
for a further 400 inner orbits. After 250~inner orbits, when the VSI
turbulence has reached a new steady state, we distribute uniformly over the
midplane between 25 and 95~au a half-million 0.1-mm and a half-million 1-mm
particles. Each particle is given the local Keplerian velocity. The Stokes number at the midplane for the 1mm grains lies between $3 \times 10^{-2}$ at the
inner boundary and $10^{-1}$ at the outer boundary. The radial midplane
profile of the Stokes number is shown in Appendix A.

\begin{figure}
\resizebox{\hsize}{!}{\includegraphics{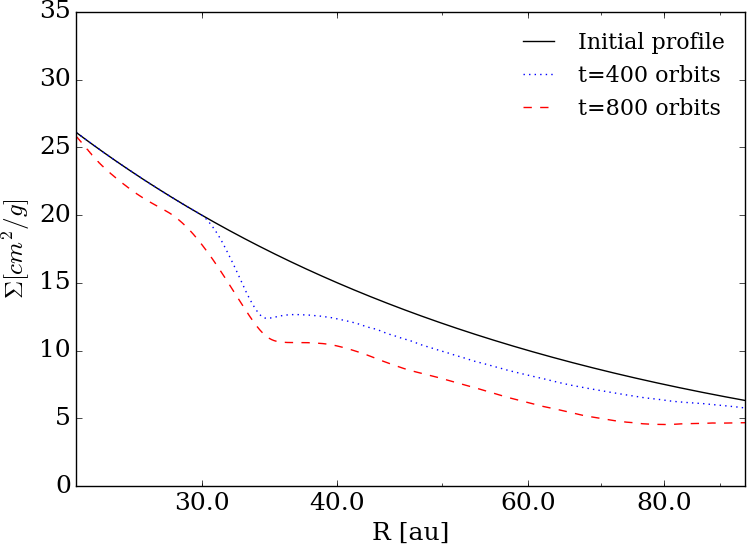}}
\caption{Radial profile of the surface density for the initial conditions at
  $t=0$ (black line), at $t=400$ orbits (blue dotted line) and $t=800$ orbits (red dashed line).} 
\label{fig:al_su}
\end{figure}

\begin{table}
\begin{tabular}{lll}
  \hline
  \hline
Surface density & $\rm \Sigma=6.0 \left ( \frac{r}{100 AU} \right )^{-1} g/cm^2$ \\
Stellar parameters & $\rm T_*=4000\, K$, $\rm R_*=2.0\, R_\sun,\, M_*=0.5\, M_\sun$\\
Opacities & $\rm \kappa_*=1300\, cm^2/g$\\
        & $\rm \kappa_d = 400\, cm^2/g$\\ 
  Dust-to-gas ratio  & $D2G=10^{-3}$ \\
  $(< 10 \mu m)$ & \\
 Grid &  $r \, : \, \theta \, : \, \phi$\\
 Domain & $\rm 20-100 AU\, : \, \pm 0.35\, rad : \, 2\pi\, rad$\\
 Resolution & $1024 \times 512 \times 2044$\\
  \hline
  \hline
\end{tabular}
\caption{Setup parameters for the 3D radiation HD disk model, including the
  gas surface density, the stellar parameters, the opacities for stellar
  irradiation and thermal emission, the dust to gas mass ratio of small grains
  ($\le 10 \mu m$), the domain and the resolution.}
\label{tab:model}
\end{table}
\begin{figure}
  \resizebox{\hsize}{!}{\includegraphics{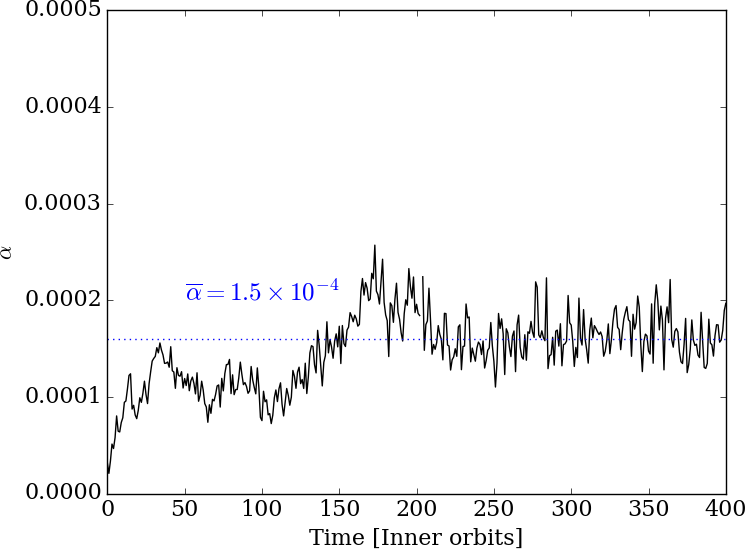}}
  \resizebox{\hsize}{!}{\includegraphics{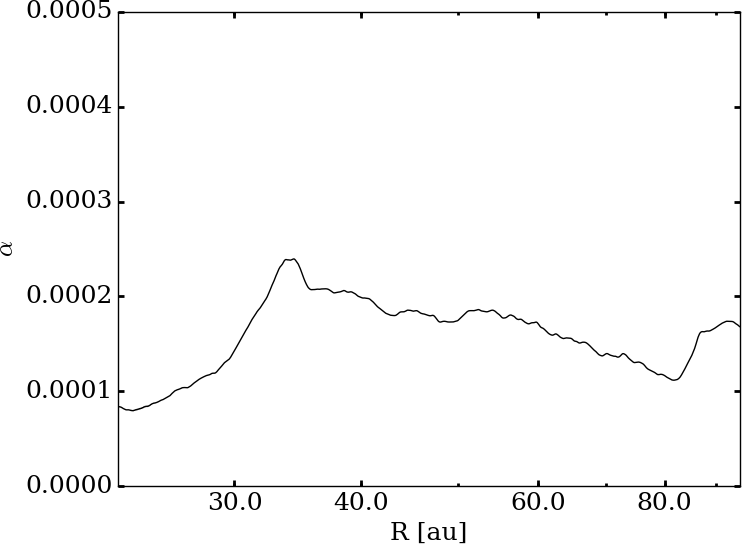}}

  \caption{{Top: Time evolution of the volume-averaged stress-to-pressure ratio
    $\alpha$. The time average between 100 and 400 inner orbits is overplotted
    (blue dotted line). Bottom: Radial profile of $\alpha$, time averaged
      between 100 and 400 inner orbits. }} 
\label{fig:alp}
\end{figure}
\section{Results}  
In the following we will first present the results on the gas kinematics, and 
describe the formation and evolution of a persistent vortex.
We will then focus on the dust evolution and the steady state dust vertical
structure that develops.
The surface density evolution is visible in Fig.~\ref{fig:al_su}, which shows the initial
condition at time zero and two snapshots corresponding to evolution after 400 and 800 inner orbits. 
The first 400 orbits were calculated with a reduced azimuthal domain, and the results for
this part of the run are presented in our previous work \citep{flo17}. The
results between 400 and 800 inner orbits were obtained after extending the azimuthal domain, 
as described in Section~\ref{sec:IC}. From now on when we discuss the results
we will let the restart time of 400 orbits correspond to time zero, as we will only discuss
the run whose azimuthal domain lies in the interval $0 \le \phi \le 2 \pi$~radians.

\subsection{Kinematics and turbulence}
\label{sec:alp}
After the full 2$\pi$~radian simulation restarts, the VSI quickly reaches a new
steady state. We gauge the turbulence level by the
stress-to-pressure ratio $\alpha$. We calculate the
pressure-weighted\footnote{The difference between the pressure-weighted and
  mass-weighted value is 0.5\%.} stress-to-pressure ratio, using
\begin{equation}
\rm \alpha = \frac{ \int T_{r \phi} dV} {\int P dV} = \frac{ \int \rho
  v'_{\phi}v'_{r}dV} {\int P dV}, 
\label{eq:ALPHA}
\end{equation}
with differential volume $dV$. The turbulent components of the velocities indicated by the superscript primes
are found by substracting the azimuthally averaged value at each r and $\theta$. The time evolution of $\alpha$ is presented in
the top panel of Fig.~\ref{fig:alp}. Over the first 30 inner orbits, the value of $\alpha$
quickly increases and saturates to a new steady level near $\alpha =1.5 \times
10^{-4}$. Because the initial state already reflects VSI turbulence, saturation takes only about one-quarter as long as in model \texttt{M3} (see Fig.~3 in \citet{flo17}). We follow the evolution for 400 inner
orbits. The fluctuations over time in the stress-to-pressure ratio
remain small compared to the mean value. 

\begin{figure}
 \resizebox{\hsize}{!}{\includegraphics{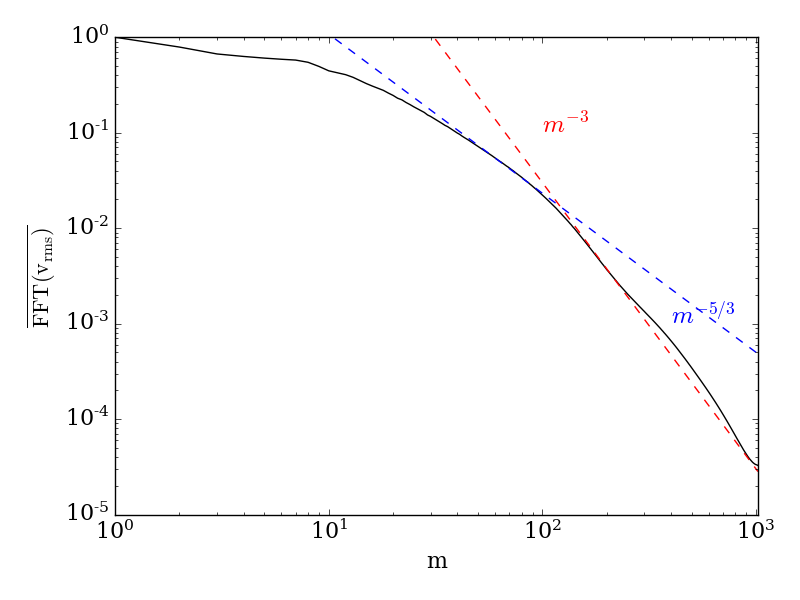}}
 \caption{Normalized azimuthal Fourier power spectrum of the RMS turbulent
   speed, averaged in space and time between 50 and 60~au, within 0.14~rad of
   the midplane, and from 100 to 400 inner orbits.} \label{fig:al_fft}
\end{figure}

The radial profile of the vertically and azimuthally averaged $\alpha$ value is shown in the bottom panel of
Fig.~\ref{fig:alp}, where an additional time average between 100 and 400
orbits has been employed. The profile of $\alpha$ shows an increase, from $8 \times 10^{-5}$ at 25 au to $2.5 \times 10^{-4}$ at 35 au. Beyond 35 au the
$\alpha$ value remains above $10^{-4}$. This increase
of $\alpha$ from 25 to 35~au is connected to the thermal relaxation timescale
relative to the threshold value for the VSI to operate. The small dip at 80 au might
be related to the run's duration. The steady state
might not be fully established near the outer radius, since the calculation ends before the outer boundary completes 36 orbits.
\begin{figure*}[t]
  \resizebox{\hsize}{!}{\includegraphics{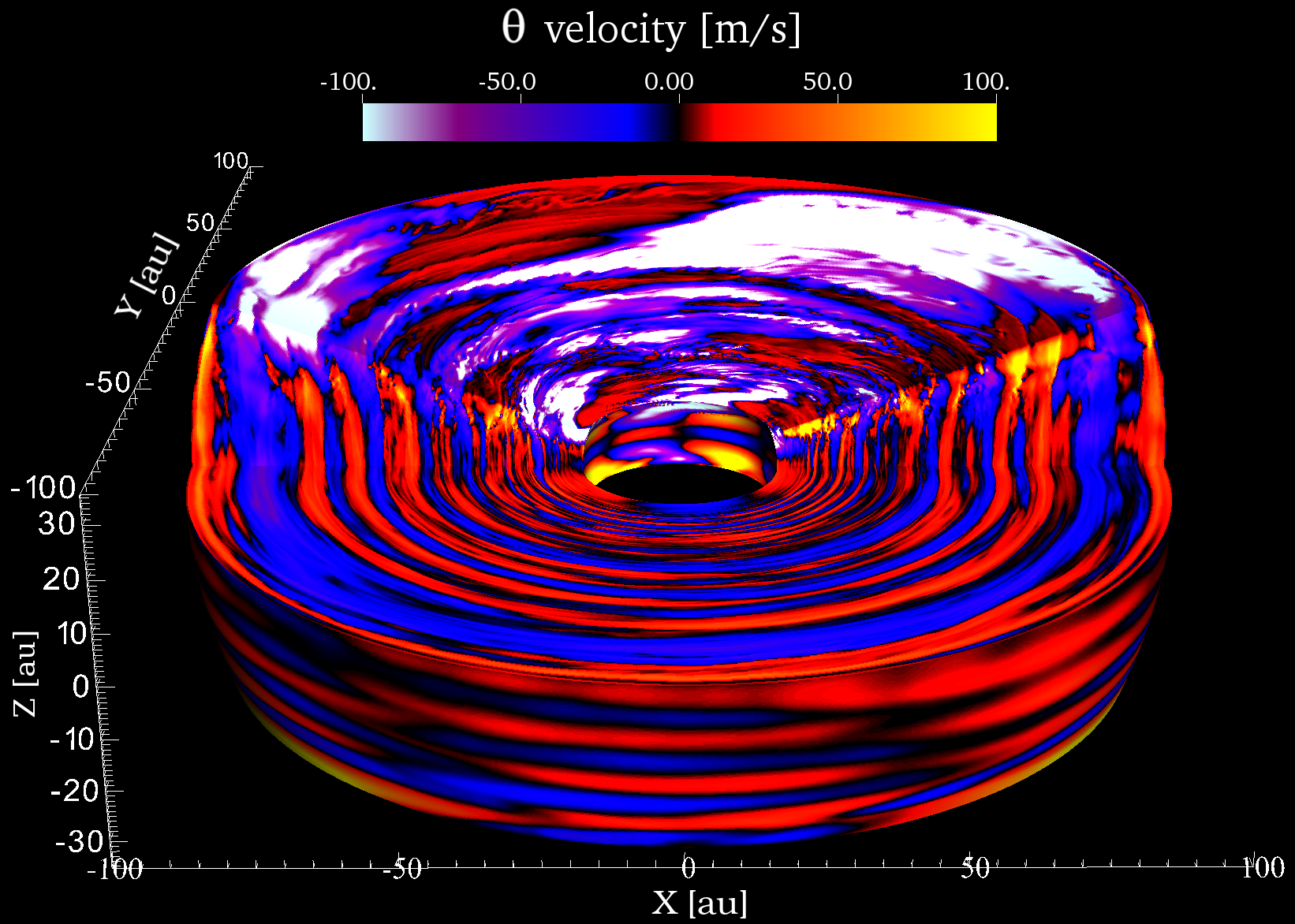}}
\caption{3D map of the $\theta$ velocity at 290 inner orbits of
    evolution.} 
\label{fig:VT}
\end{figure*}

The turbulent flow speed $v_{rms} = \left\{(v^\prime_r)^2 +
  (v^\prime_\theta)^2 + (v^\prime_\phi)^2\right\}^{1/2}$ near the midplane
well outside the 35~au surface density dip has the azimuthal power spectrum
shown in Fig.~\ref{fig:al_fft}.  This spectrum is averaged over the volume
between 50 and 60~au in radius and within 0.14~radians of the midplane in
$\theta$.  The spectrum is further time-averaged from 100 to 400~inner orbits,
and normalized by its maximum value.  The high spatial resolution means we
cover three decades in azimuthal wavenumber. The turbulent spectral energy
peaks at the scale of the circumference, with a shallow decline towards
shorter scales.  For azimuthal mode numbers $m$ of 10 to 100 the slope turns
steeper, passing through $m^{-5/3}$.  Between $m = 100$ and $1000$ the slope
turns steeper again, and the turbulent motions at the smallest resolved scales
carry four orders of magnitude less spectral power than those at the largest
azimuthal scales. \citet{man18} found a similar spectral shape in their VSI
turbulence models, though the steepening seen here at $m=100$ occurs in their
models at $m=40$, likely due to the lower numerical resolution, steeper radial
pressure profile, or differing thermal relaxation timescale. \citet{man18}
used a uniform value $t_{relax} = 2 \times 10^{-3} / \Omega$, comparable to our smallest value here on the midplane.

Fig.~\ref{fig:VT} shows a snapshot of the total $\theta$-velocity at 290~inner orbits.  Cuts along the midplane and a representative meridional plane show the 3D structure.  The characteristic pattern of the VSI is visible, with large-scale upward and downward motions crossing the midplane and reaching the disk atmosphere.  Speeds are as great as 100~m~s$^{-1}$ in the atmosphere and the midplane motions are almost as fast, with amplitudes up to 50~m~s$^{-1}$.  The midplane cut makes clear that the motions are approximately axisymmetric, with small but noticeable deviations especially at low wavenumbers.
\begin{figure*}[t]
  \resizebox{\hsize}{!}{\includegraphics{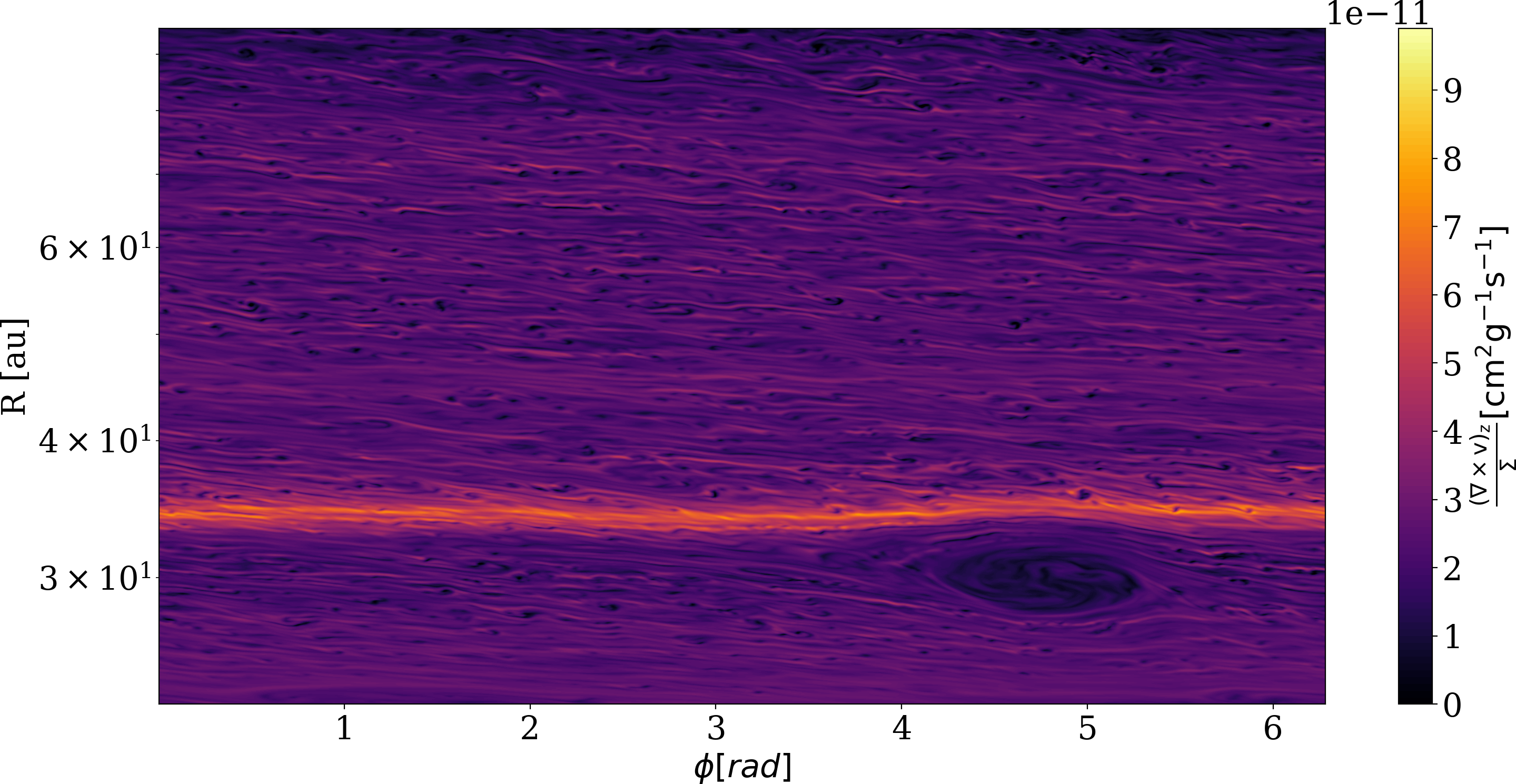}}
\caption{Midplane vortensity at 300 inner orbits. A large vortex, RWI
  generated, is seen at 30~au.} 
\label{fig:vort}
\end{figure*}

\begin{figure}
  \resizebox{\hsize}{!}{\includegraphics{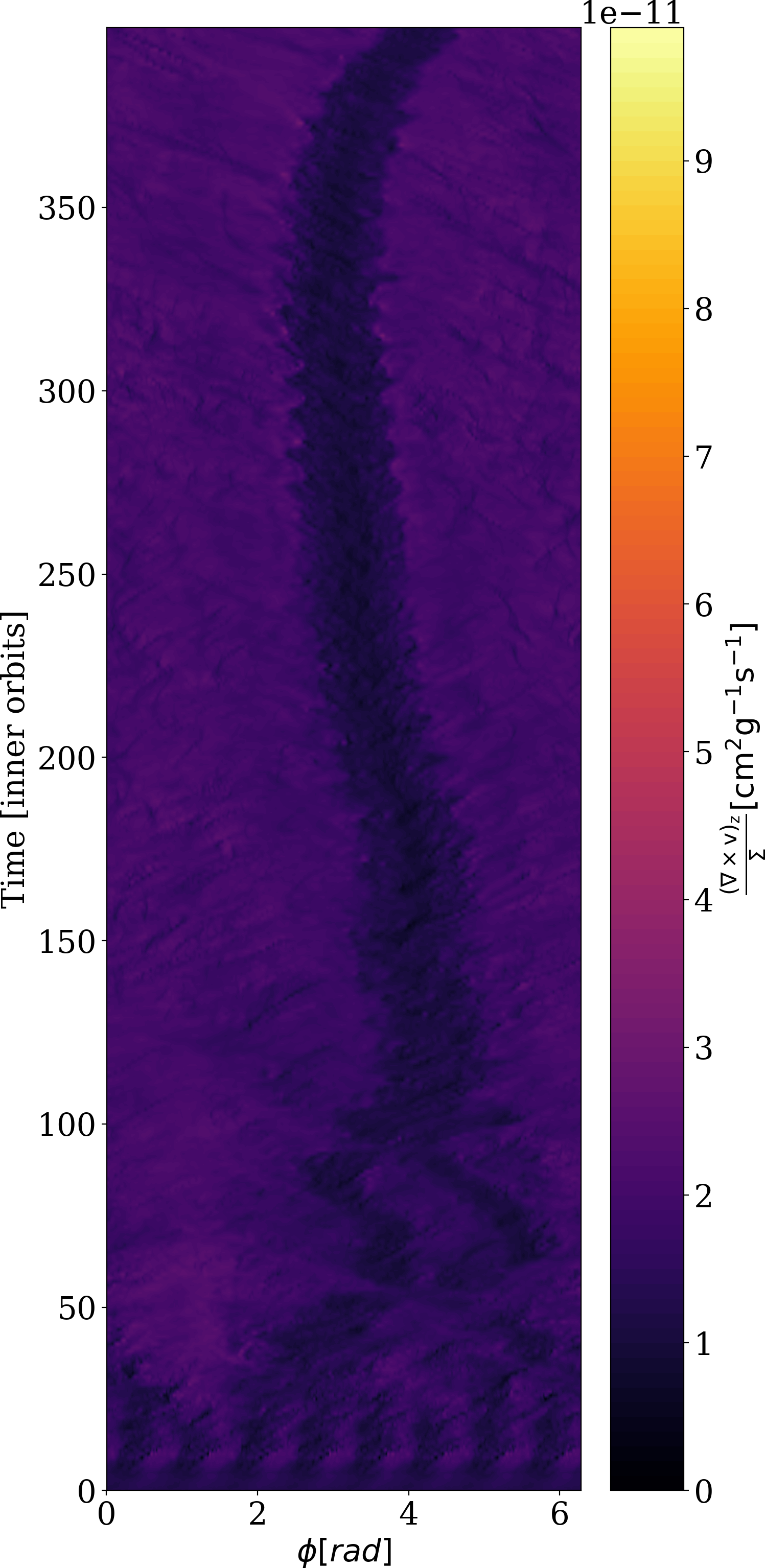}}
  \caption{Vortensity versus azimuth and time calculated at the
  midplane of the disk and averaged between 29 and 31 au. After vortices grow
  by RWI and merge up through 100 orbits, the merger remnant remains similar
  in size and strength for 300 inner orbits ($\sim 163$ local orbits).} 
\label{fig:vort_time}
\end{figure}

\subsection{Vortex formation and evolution}
\label{sec:Vortex}
Here we investigate the formation and evolution of vortices in the simulation. 
We identify vortices using the vertical component of the vortensity
$\frac{(\nabla \times v)_z}{\Sigma}$. Fig.~\ref{fig:vort} shows a map of the vortensity in the $r-\phi$ midplane after 300 inner orbits. A single large vortex is visible as a dark
region with a minimum in the vortensity around 30 au. The
vortex is located radially inwards from the gas gap which is visible as a bright ring.
The vortex is fully developed after 150 inner orbits. The position
and size of the vortex are listed in Table~\ref{tab:vor}. We define the vortex
as the region where $\frac{(\nabla \times v)_z}{\Sigma} < 10^{-11} cm^2 g^{-1} s^{-1}$. The radial extent fluctuates between $\rm 3.6$ and $\rm 4.6au$, the azimuthal
extent between  $\rm 35 $ and $\rm 45au$, and the aspect ratio between $\chi =
8.7$ and $\chi = 12.6$. This vortex is generated by Rossby-wave instability
and occurs in an annulus where the instability criterion of \citet{li00} is
satisfied as we show in Appendix~B. The detailed time evolution of the vortensity at 30 au is
plotted in Fig.~\ref{fig:vort_time}. During the first 50~inner orbits, the
eight-fold symmetry from replicating run M3 is broken and a new steady-state
develops.  Small vortices form and merge up to 100~inner orbits.  Thereafter a
single large vortex is present.  To lessen this vortex's drift sideways on the
plot, we shift each row in azimuth according to the orbital speed at
30~au. The vortex migrates only very slowly in radius (Table~\ref{tab:vor}). A number of factors influence the speed at which a
vortex migrates, including its size, strength, and aspect ratio \citep{paa10}. For aspect ratios $\chi \ge 7$, \citet{paa10} showed that vortex
migration can be very slow, being two orders of magnitude slower than for a
vortex with aspect ratio $\chi = 2$, for example, and hence it is likely the large aspect ratio of the vortex formed in the simulation ($\chi \sim 10$)
that explains its slow migration. Also important to note is that the vortex does not decay and is present for
the whole remaining duration of the simulation. In contrast to our previous models of MHD turbulence 
arising from the magnetorotational instability (MRI),
where a vortex was repeatedly formed and destroyed at the MRI dead-zone outer edge
\citep{flo15}, here the vortex is not destroyed. Ignoring MHD effects,  \citet{les09} showed that elliptical vortices are
unstable with the exception of vortices with an aspect ratio between 4 and 6,
which are stable. However the growth rate of the elliptical instability which
could destroy the vortex remains small. For the aspect ratio of the large
vortex we predict a growth time of about 100 orbits at the vortex position \citep{les09}.
Also, as this vortex is sustained by the RWI it might not be subject to the elliptical instability \citep{lyr13}. Further longer-term studies are needed
to test the stability of this vortex, but close inspection of Fig.~\ref{fig:vort} indicates that the flow inside the vortex is not smooth, suggesting that the elliptical instability is operating to generate turbulence inside the vortex, but the tendency of this interior flow to destroy the vortex is counteracted by the continuous driving provided by the RWI due to the persistent pressure bump.

Many small, short-lived vortices present in our simulation are visible in
Fig.~\ref{fig:vort}. These appear similar to those
reported by \citet{ric16}, and likely have the same origin, namely small scale
vorticity/vortensity perturbations generated by the VSI that undergo, local
Rossby wave instability. In Appendix C, we isolate and observe the formation and
evolution of such a small vortensity minimum. The lifetime is found to be approximately 1.4
local orbits, and this short life time is likely due to the elliptical instability acting strongly in these vortices which have small aspect ratios. As discussed in the next section, small local dust concentrations arise in these vortices, but are limited by the short lifetimes.

\begin{table}
\begin{tabular}{lllll}
time & $r$ [au] &  $\Delta r$ [au] & $r \Delta \phi$ [au] & $\chi$ \\
\hline
150 &30.1 & 4.6 & 45.1 & 9.8  \\
200 & 30.4 & 3.6 & 45.3 & 12.6 \\
300 & 29.9 & 4.1 & 35.6 & 8.7  \\
400 & 29.5 & 3.6 & 34.8 & 9.6  \\
\hline
\end{tabular}
\caption{Vortex parameters: time in inner orbits, radial position, radial width, azimuthal extent, and aspect ratio.}
\label{tab:vor}
\end{table}

\begin{figure*}[t]
  \resizebox{\hsize}{!}{\includegraphics{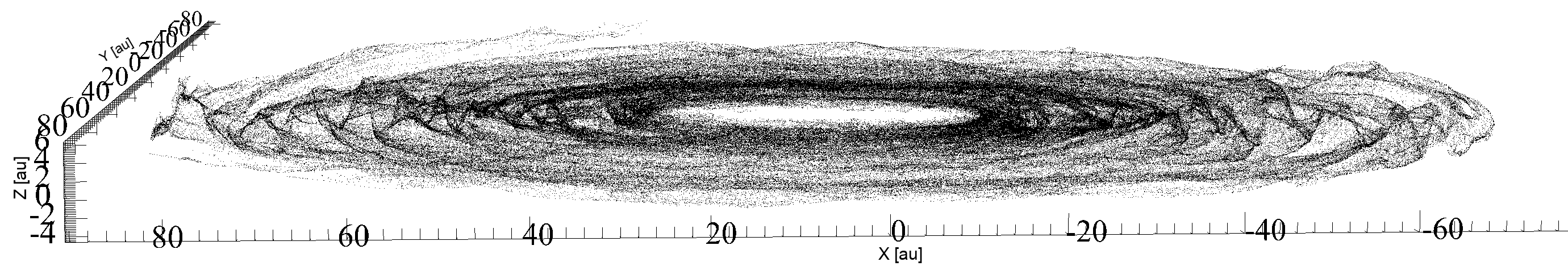}}
  \caption{3D plot of the 1mm grains (black dots) viewed nearly edge-on, at 50
    inner orbits after injection. The grains' vertical extent remains small. } 
\label{fig:edge}
\end{figure*}

\begin{figure}
  \resizebox{\hsize}{!}{\includegraphics{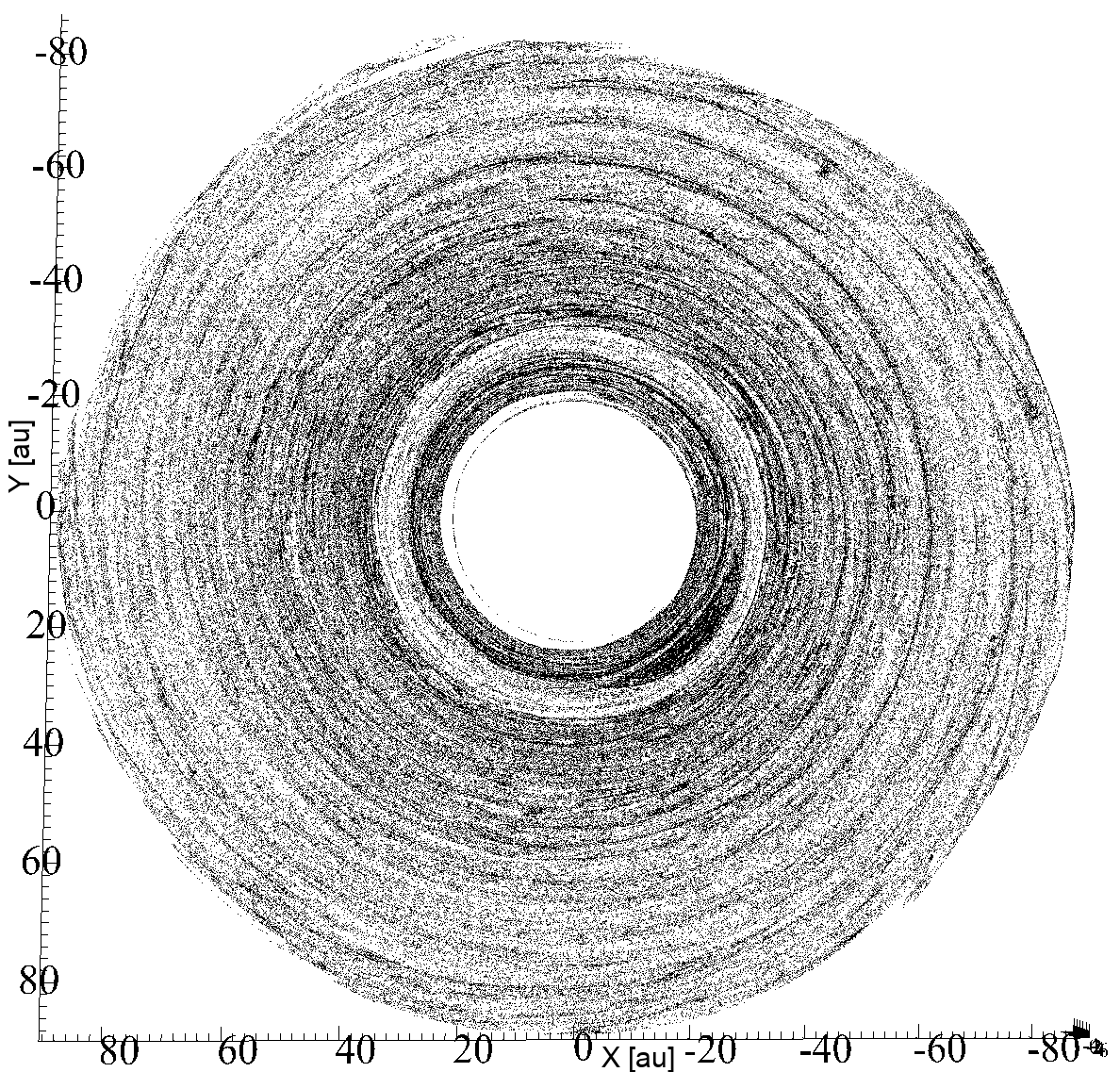}}
\caption{3D plot of the 1mm grains (black dots) viewed face-on, at the same
  time as Fig.~\ref{fig:edge}. Grains are concentrated inside the vortex at the four o'clock position.} 
\label{fig:face}
\end{figure}

\subsection{Dust evolution}
In this section we describe the results from the evolution of the 100 $\rm \mu m$
and 1 mm grains. A snapshot of the 1mm grains 50~inner orbits after the grains
were added is shown in
Figs.~\ref{fig:edge} and \ref{fig:face}, close to edge-on and face-on,
respectively. Most of the 1mm grains remain within $\rm \pm 4 au$ of
the midplane. In the outer regions, an oscillatory
structure becomes visible due to the combination of radial drift and the
vertical motions induced by the VSI. We note that in the outer disk the grains are not
yet fully vertically mixed. For the 1mm grains we measure a mean inwards drift speed
of $\rm 7.5\, m/s$ outside of 70 au, which roughly corresponds to 0.2 au per
inner orbit. This value is only slightly less than the analytical estimate of $\rm 9\, m/s$ using the initial
profiles. Fig.~\ref{fig:face} shows the distribution in a face-on projection at the
same time. The distribution of 1mm grains clearly shows the effect of
the surface density depression with a lack of particles in this region. The vortex at the inner
edge of the steep drop is clearly visible as a concentration of the 1mm grains
to the south-east.
The other disk regions exhibit small scale concentrations of particles. These
concentrations are connected to the short lived vortices described in Section~\ref{sec:Vortex} and
\citet{ric16}. A fully edge-on view and comparison between both grain sizes can be
found in Appendix D.

\begin{figure}
  \resizebox{\hsize}{!}{\includegraphics{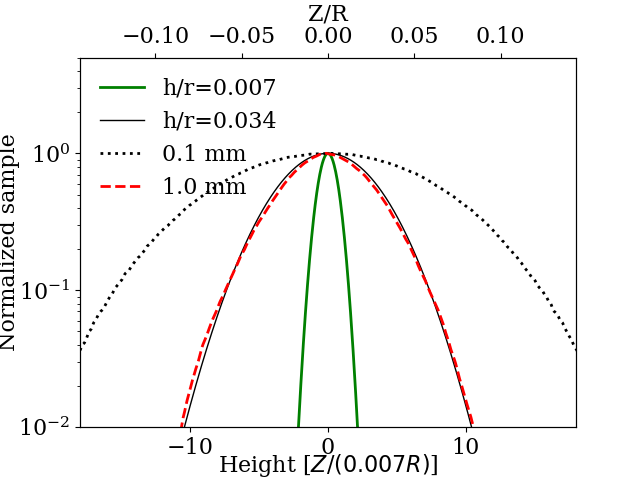}}
  \caption{{Normalized vertical distribution for 0.1 and 1 mm grain sizes for
      an annulus between 30 and 40~au, time and space averaged. The green line shows the best-fit value
      by \citet{pin16} for HL Tau with $H/R=0.007$.}} 
\label{fig:dust_ver}
\end{figure}
\subsection{Dust vertical structure}
A critical ingredient for comparing 3D simulations of gas and dust with
observations is the scale height of the dust grains. To
measure the vertical distribution of the grains we consider all particles in
the radial interval between 30 and 40 au. The particles are sampled by binning
onto a
high resolution grid with 1000 cells in the $\theta$ direction. The
distribution is calculated at each inner orbit and then averaged from 30 inner
orbits after introducing the dust particles until the end of the run. As we have shown in \citet{flo17}, grains at this position have
already been mixed well enough to describe a new steady profile. The results are shown in Fig.~\ref{fig:dust_ver}.
The plot shows that the 1 mm grains are mixed up very efficiently.
We fit the dust scale height $h_d=H_d/R$ with a value of $h_d^{1mm}=0.034$. This value is much higher than has been inferred from observations of 
the HL Tau system \citep{pin16}, with $h_d^{1mm}=0.007$, and higher than what we found
in our previous non-ideal magnetohydrodynamical models \citep{flo17}.

We also compare our results with the well-known equilibrium solution for dust
settling in a turbulent medium \citep{dub95}
\begin{equation}
h_d=\frac{h_g}{\sqrt{1+ \frac{\mathit{\rm St}}{\widetilde{D}} }}
\label{eq:hd}
\end{equation}
with the gas scale height in units of the radius $h_g=H/R$, the Stokes
number, $\mathit{\rm St}$, and the dimensionless diffusion coefficient
$\widetilde{D}$ defined as $D=\widetilde{D}c_sH$. The dimensionless diffusion
coefficient is often expressed in units of the $\alpha$ parameter
$\widetilde{D} = \alpha/\mathit{\rm Sc}$ \citep{dul04,sch04} with the Schmidt
number $\mathit{\rm Sc}$ close to unity \citep{joh05,tur06}. %stress-to-pressure ratio, $\alpha_z$ which
By using the values from our simulations for the 1 mm grains between 30 and 40
au, $\mathit{\rm St} =5 \times 10^{-2}, \alpha = 2 \times 10^{-4}$ we obtain $h_d=0.007$, a similar value to what we
found in our non-ideal MHD simulations \citep{flo17a} and to what has been inferred for the mm dust
in HL Tau \citep{pin16}. However to obtain the value $h_d=0.034$, which we directly fitted to the dust
grain distribution, we need to assume $\alpha_z=5.4 \times 10^{-3}$, which tells us that
the VSI is 27 times more efficient in mixing vertically than in transporting
angular momentum radially. We therefore suggest that if the VSI is operating in the HL Tau disk, the
vertical mixing might be quenched either by non-ideal MHD effects by a
magnetized wind layer \citep{cui19} or by dust drag due to a high dust to gas
mass ratio \citep{pin16,lin19}. Recently \citet{sch20} obtained results
showing the VSI operating together with a dust layer. The efficiency of the
streaming instability in these environments remains still under investigation
\citep{yan18,umu19,sch20,che20}. We discuss more on the dust scale height measurements in Section 4.7.

Following \citet{fro06b}, another way to express the vertical diffusion is
given by $D=\tau_{eddy} \left <v_z^2 \right >$.
To compare with previous models by \citet{sto16} we determine the eddy
timescales $\tau_{eddy}$. For the 1mm grains we determine in the annulus
between 30 and 40 au $(<v_z^2>)^{1/2} = 0.0166 c_s$ which leads to
$\tau_{eddy} = 0.209 \Omega^{-1} $. A similar value $\tau_{eddy}\approx 0.2$ was found by \citet{sto16} for
the vertical diffusion of particles (see Section 4.3 therein).
\subsection{Individual grain motion}
\begin{figure}
  \resizebox{\hsize}{!}{\includegraphics{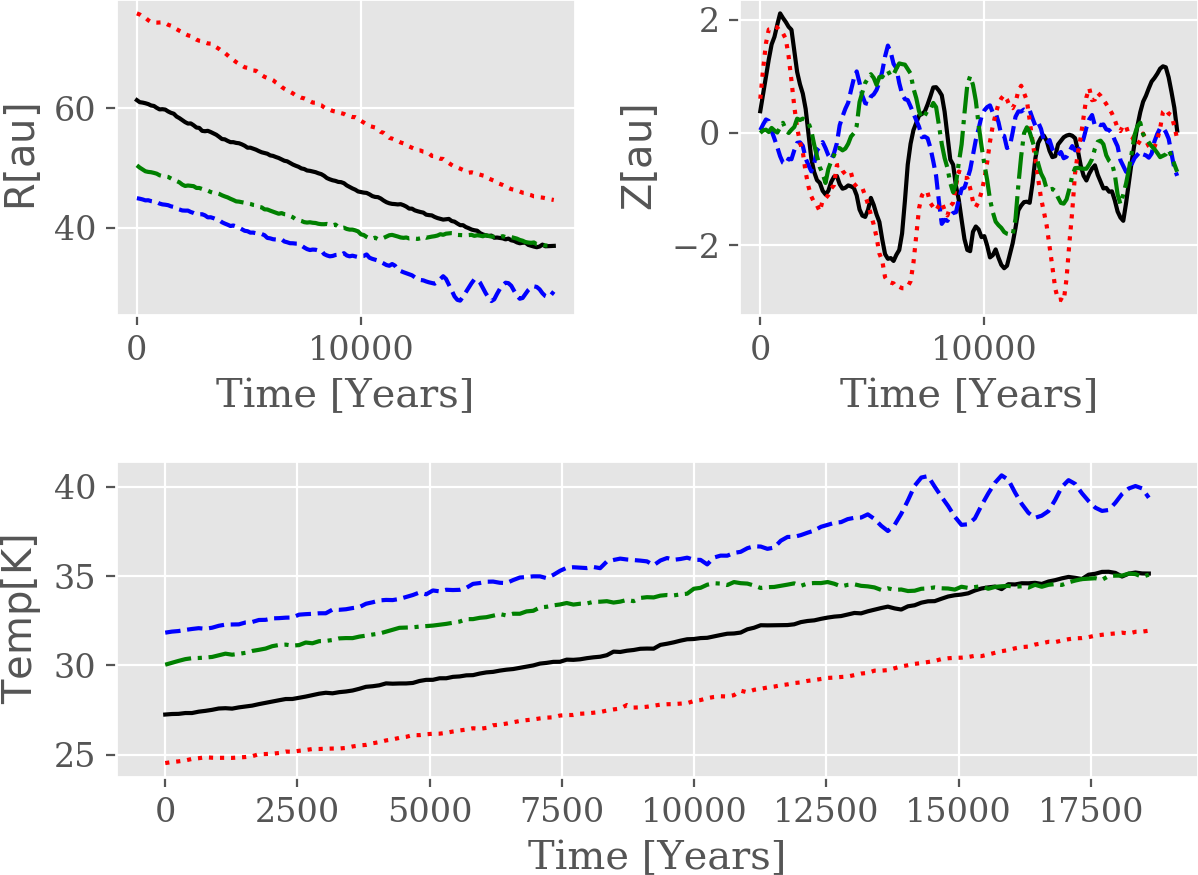}}
  \caption{Spatial and thermal evolution of four individual 1-mm grains over time.} 
\label{fig:dust_1mm}
\end{figure}

We illustrate how the grains evolve by here showing four representative
particles, chosen at random from among the 0.1-mm particles, and four from
among the 1-mm particles. Fig.~\ref{fig:dust_1mm} shows the spatial and thermal evolution of the four
1-mm grains.
All four show a clear radial drift over the simulation runtime. The
time is now given in years as their radial location is constantly changing. The 1mm grains drift inwards approximately 20~au in 10,000 years. Fig.~\ref{fig:dust_1mm}, top left panel, also shows that
locally the radial drift can be reduced (see green dash dotted line) or
can halt inside the vortex (blue dashed line). The vertical
distribution of the four grains is shown in Fig.~\ref{fig:dust_1mm}, top
right panel. The VSI's characteristic upward and downward motions are clearly
visible. At these distances, mm grains need around 5000 to 10000 years to
travel from one hemisphere to the other, around 2 au above and below the midplane.
None of the four mm grains make excursions up into the regions of the disc that are strongly heated by the star, as is clear from their temperature evolution. Fig.~\ref{fig:dust_1mm}, bottom panel, shows the temperature
evolution of the grains. We assume the dust and gas temperatures to be equal.
All four grains warm up as they drift inward. The grain
which gets trapped in the vortex (blue dashed line) shows oscillations in
temperature around 40~K, connected to the radial oscillations due to the vortex
rotation.
\begin{figure}
  \resizebox{\hsize}{!}{\includegraphics{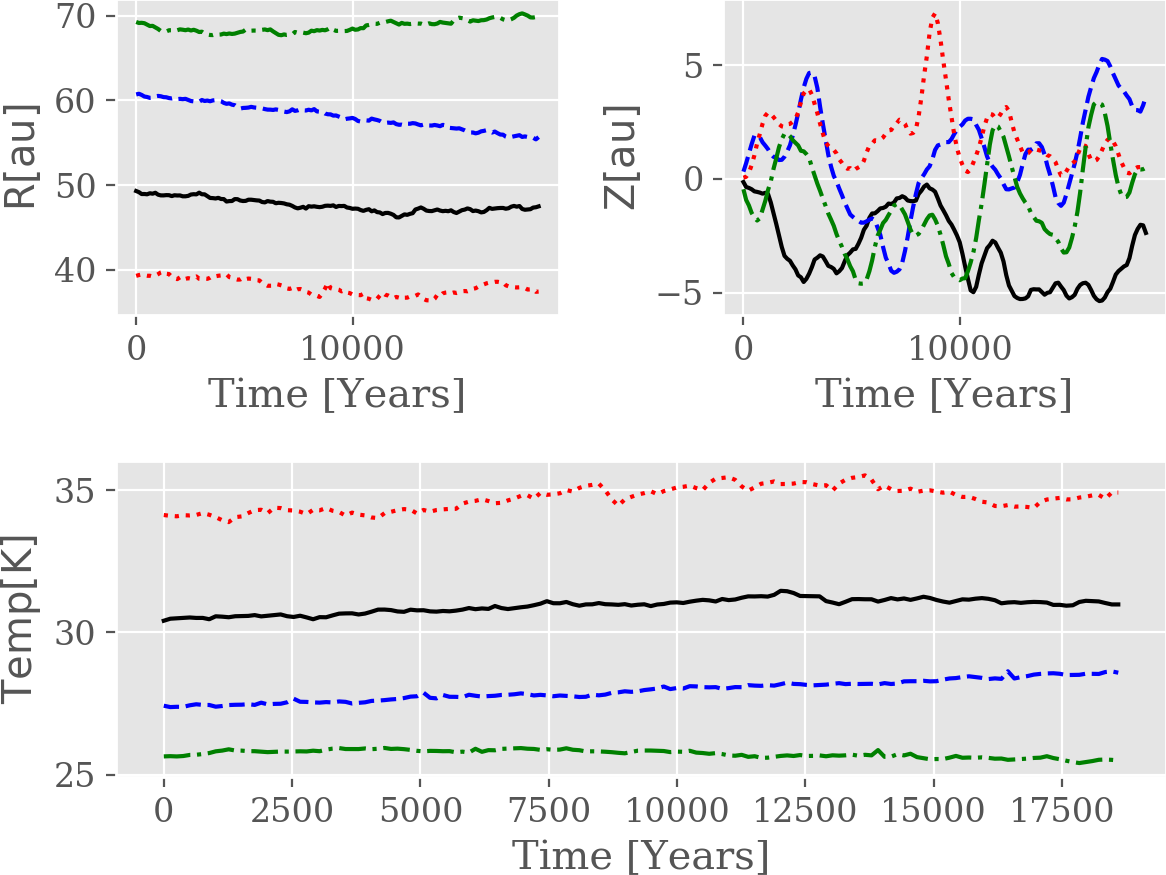}}
\caption{{Spatial and thermal evolution of four individual 0.1-mm grains over time.}} 
\label{fig:dust_01mm}
\end{figure}
The evolution of four individual 100 $\mu m$ grains is shown in
Fig.~\ref{fig:dust_01mm}. The top left panel shows the
radial locations over time. As expected the radial drift of these grains is
strongly reduced compared to the 1 mm grains. Three of the grains show no substantial
radial drift, while one drifts about 5 au in 20 thousand years (blue dashed line).
Fig.~\ref{fig:dust_01mm}, top right panel, shows the vertical locations of the grains
over time. Due to the their smaller size they are able to reach higher
altitudes in the disk, spanning a region $\pm 5 au$ about the midplane,
showing a similar pattern of upward and downward motions to the 1-mm grains.
The 100~$\mu$m grains also remain below the disk's irradiated surface layers,
as shown by the temperature evolution in Fig.~\ref{fig:dust_01mm} bottom
panel. The warm starlight-irradiated layers begin about 2.2~scale heights or 12~au above and below the midplane at 50~au.
\subsection{Turbulent speeds}
\begin{figure}
  \resizebox{\hsize}{!}{\includegraphics{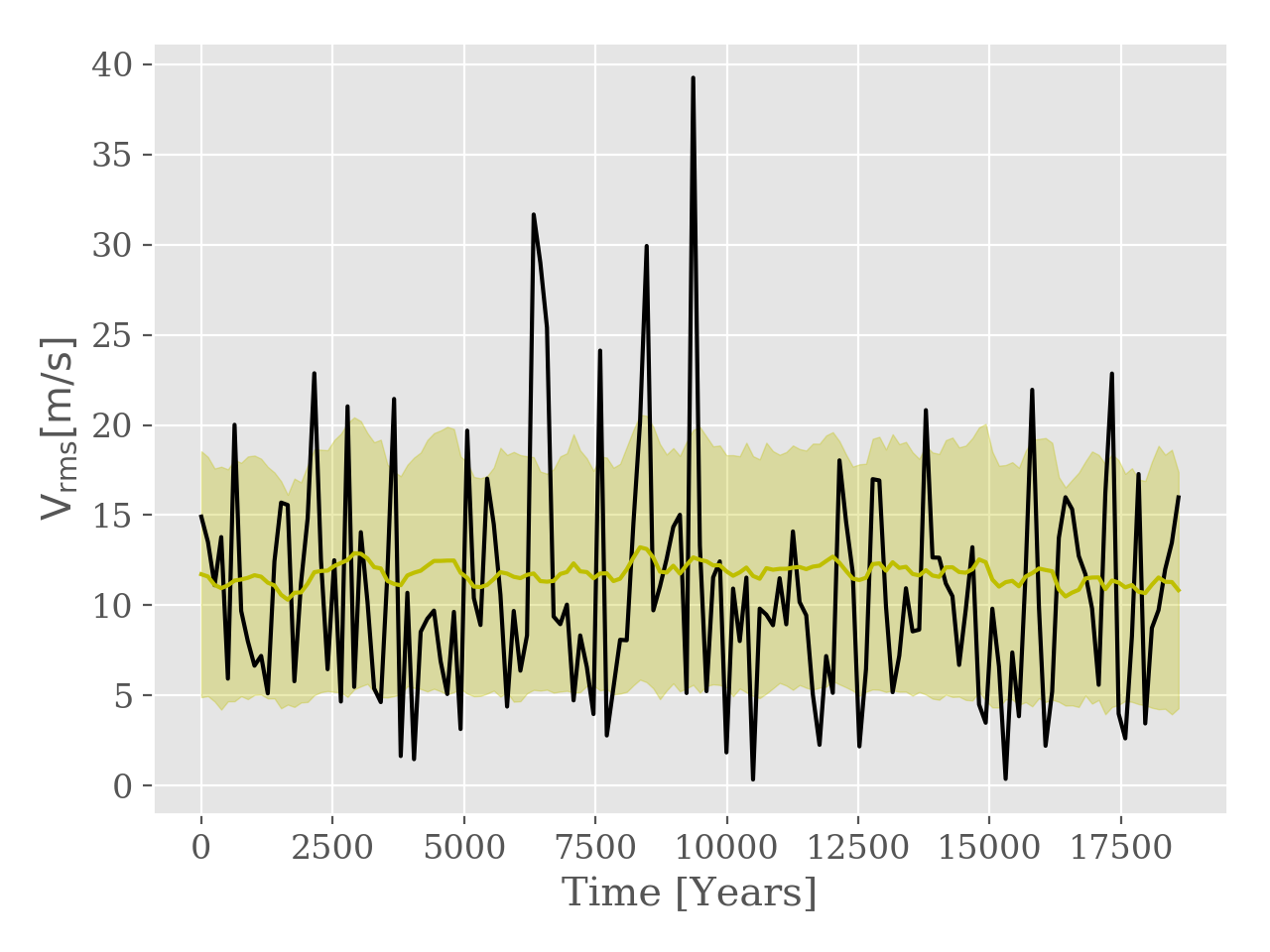}}
\caption{Turbulent speed of one individual 1-mm grain (black solid
    line) and the average of all 1mm grains (yellow solid line) plus and minus one standard deviation (yellow shading).} 
\label{fig:dust_vel}
\end{figure}

To obtain the characteristic turbulent velocities of the grains we first
perform a time average over the full simulation time of each velocity component. This temporal mean is
subtracted from each particle's velocity at each time step. This
way the radial drift is removed from the radial component. The azimuthal
component changes erratically with time, such that a mean azimuthal velocity
is not easily chosen. We therefore focus here on the radial and vertical turbulent
velocity components, and note that these provide a lower limit on the turbulent flows' speed.
Fig.~\ref{fig:dust_vel} shows the time evolution of the turbulent speed for an individual 1mm grain. The root-mean-square and standard deviation
of all remaining 1mm grains are over-plotted for reference.
We determine a value of $\rm v_{RMS}^{r,\theta}=11.7 \pm 6.7 m/s$. The 100 $\mu m$ grains show a slightly
higher turbulent of $14.3 \pm 8.0 m/s$. The representative individual particle
plotted in Fig.~\ref{fig:dust_vel} shows that
locally grains are accelerated quite substantially to several tens of
meters per second. For the chosen mm grain, for a short time period the speed
exceeds 30~m/s. 
\section{Discussion}
In this section we compare our results with other models and
discuss the implications of our findings. We start with the turbulent
saturation value, followed by the radial profile of $\alpha$, then vortex
formation, and finally the dust mixing.
\subsection{Turbulence saturation value}
With our high resolution simulations covering all $360^\circ$ of azimuth we reach a turbulence level of $\alpha \cong
10^{-4}$. We note that this value is around 4 times higher than what
we have found for our previous model with a limited azimuthal extent. This
result is not surprising as larger modes can now fit into the domain and it is known that the VSI
drives perturbations with large azimuthal scales (see also the spectrum
analysis in Fig.~\ref{fig:al_fft}).
Our value of $\alpha$ is similar to previous radiation HD simulations covering
$90^\circ$ in azimuth by \citet{sto14} which also found  $\alpha
\cong 10^{-4}$. Isothermal simulations by \citet{nel13} and \citet{man18} reported higher
values with $\alpha \cong 10^{-3}$. Such higher values are expected for
isothermal simulations as those present the upper limit of fastest cooling
times. In addition we note that the VSI strength depends on the disk's aspect
ratio \citep{nel13} with stronger turbulence for larger scale heights. Our
aspect ratio H/R ranges from 0.09 at the inner radial boundary to 0.14 at the outer boundary.
We also highlight the different nature of VSI turbulence compared with the
turbulence driven by magneto-rotational instability (MRI). %The
The azimuthal Fourier power spectrum we obtain for the VSI is similar to
previous results. The slope is shallow between $m=1$ and
$m=10$, steepening to $m^{-5/3}$ between $m=10$ and $m=100$, and steepening
further between $m=100$ and $m=1000$. \citet{man18} reported a similar profile
for wavenumbers up to $400$.
\subsection{The inner edge of the VSI active zone}
\begin{figure}
  \resizebox{\hsize}{!}{\includegraphics{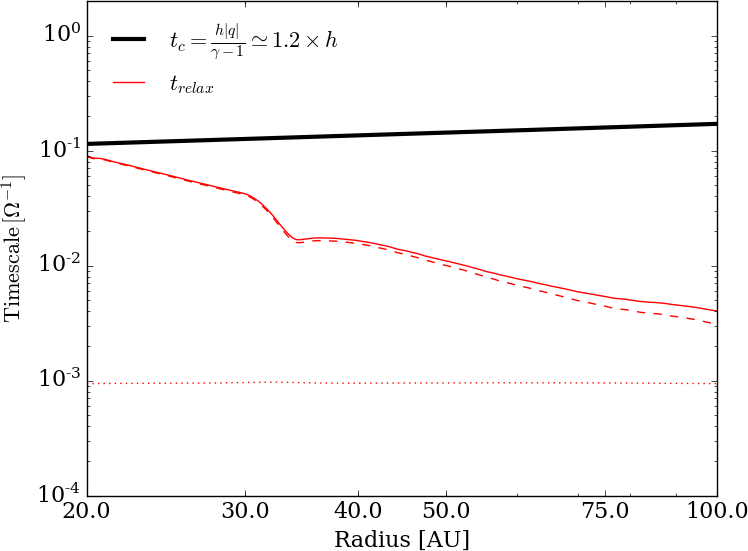}}
\caption{Radial profiles of the critical timescale for the VSI (black) and
the relaxation timescales at the midplane (red solid curve) for the time of restart after
400 inner orbits with $q=-0.5$ as the radial temperature profile exponent. The red dotted and dashed lines show the contribution from
the optically thin and thick timescales, see \citet{flo17}. The vertical axis is
in units of the local inverse Keplerian frequency.} 
\label{fig:tc}
\end{figure}

Our simulation results show increasing VSI activity with radius out to 
35 au, after which the strength of the induced turbulence levels off. The
increase is connected to the fact that the relaxation timescale is a
decreasing fraction of the orbital period as we move away from the star,
falling further below the threshold value below which the VSI operates. Moving
radially outwards, the disk becomes less optically-thick, allowing faster
thermal relaxation and strengthening the VSI.
{ \citet{nel13} found the threshold for the relaxation timescale to be $t_{relax} = 0.1
  \Omega^{-1}$. For $t_{relax} = 0.1$ they found that the VSI does not operate
  in the disk while decreasing the relaxation time by only one order of magnitude was
  enough to show a fully evolved turbulence (compare Fig. 12 in
  \citet{nel13}).} In Fig.~\ref{fig:tc}, we compare the
critical timescale of the VSI with the thermal relaxation timescale as
described  in \citet{lin15} and adopted by \citet{mal17} and
\citet{lyr19}. For our model we expect the VSI to be quenched inwards of 20
au. We find that between 20 and 30 au the turbulence strengthens, saturating at around 35 au (see Fig.~\ref{fig:alp}).
This turbulence profile triggers a modification of the surface
density profile, and the structure that develops 
leads to the subsequent generation of a vortex by the RWI.
It is interesting to note that such a process of surface density perturbation is very similar to the vortex
generation mechanism in a magnetised disk which has a dead-zone in the inner disk but which supports 
magnetorotational turbulence in the outer disc. Here the change in turbulence activity also leads to
the formation of a vortex at the outer edge of the dead-zone \citep{flo15}. We interpret our new result
of a large scale and persistent vortex forming in a non-magnetised disk
in terms of a VSI dead-zone transitioning to an active zone once the thermal equilibrium
timescale reaches the critical timescale of the VSI. { We note that the
  exact location and size of the vortex might be influenced by the initial surface density profile from which we restarted. We expect that the VSI transition region should also occur in simulations having a pure $\beta$
  cooling which depends on radius and height. 2D maps of the thermal
  relaxation timescale were presented by \citet{pfe19}, and these
  could easily be implemented within the $\beta$ cooling framework. Future studies should test this transition regime of the VSI and its possible impact on ring formation in protoplanetary disks.}  

\subsection{Dust opacity feedback on the VSI activity}
The position of the inner edge of VSI activity depends on the optical depth and so on
the total amount of dust. For a lower dust abundance, the VSI activity is
shifted inwards \citep{flo17} while for a higher dust amount it is shifted
outward. For our model we estimate the inner edge of the VSI activity to be
at about 45 au, 18 au and 7 au for dust-to-gas mass ratios of small grains
of $10^{-2}$, $10^{-3}$ and $10^{-4}$. 

The fact that the VSI activity is sensitive to the total optical depth and so the
local dust abundance suggests positive feedback could occur. For example, a
small local random fluctuation in the accretion stress may lead to the
formation of neighbouring disk annuli with higher and lower surface densities. The resulting changes in the local
optical depths might reinforce the variation in accretion
stress, allowing the changes to amplify. Such a process may
lead to the formation of ring-like perturbations in the surface density
profile, as demonstrated by \citet{dul18a}. Furthermore, substantial concentration of grains may have a positive dynamical feedback effect, as the additional inertia the grains provide might further reduce the turbulent activity in the higher surface density regions relative to the surrounding lower density regions \citep{lin19}.

\subsection{Vortex formation}
Our model develops a large-scale, long-lived vortex at around 30 au.
This vortex appears because of the surface density perturbation arising from
the spatial variation in turbulent activity. The vortex aspect ratio of 
$\chi \sim 8$ is very similar to the value reported in
\citet{man18}, who also saw large-scale, long-lived vortices 
in their simulations of the VSI operating in discs with very short cooling timescales. 
Unlike their vortex, the one in our model does not migrate because it is 
generated and maintained at a single persistent feature in the surface density profile. The vortices in \citet{man18} appear to be generated by the RWI operating at the locations of
intermittent surface density features that arise as natural fluctuations in the flow.  We do not
see the spontaneous emergence of large scale vortices by this mechanism in our simulation, likely due to the fact that the cooling timescales adopted by \citet{man18} lead to a more vigorously
operating VSI with a larger $\alpha$ value of $\sim 10^{-3}$, compared to the smaller value of
$10^{-4}$ we obtain in our simulations. Fluctuations around this larger value of $\alpha$ likely lead to
larger fluctuations in the surface density profile, allowing the emergence of
multiple large scale vortices. \citet{man18} also use steeper power-law radial
declines in density and temperature, strengthening VSI activity and vortex formation.

In addition to the large persistent vortex, we observe the formation of many smaller vortices
that have short life times, on the order of the orbital period, because of their small aspect ratios. 
Both \citet{ric16} and \citet{man18} have reported the existence of similar small and short lived vortices
in simulations of the VSI. These could arise from secondary instability
because of the RWI or Kelvin-Helmholtz instability which acts on small scale vortensity perturbations
generated by the VSI \citep{lat18}. The formation of small, tightly wrapped
vortices with small aspect ratios allows the elliptical instability to operate
effectively, such that the vortices quickly dissolve into the background
turbulent flow. { Vorticity capturing schemes, such as that presented by \citet{sel17}, might be well suited for studying such small scale vorticies. Finally, we note that the reduced resolution of 35 cells per H in azimuth should nonetheless be high enough to study the formation and evolution of vortices. For comparison \citet{man18} used a resolution of 12 cells per H.}

\subsection{Dust mixing and depletion of CO}
In a recent work \citet{kri18} found that dust growth and settling can lead to a
depletion of CO from the warm molecular layer located at higher altitudes in the disk. 
Connected to this idea we
investigated whether the strong upward motions of the VSI can lift 100 $\mu m$
sized grains into these warm molecular layers. However, none of the
half-million 100 $\mu m$ grains in our models reaches these irradiated
layers. Nonetheless, the idea that grains in the cold midplane can be mixed up
into the warm molecular layer, and release molecules to the gas phase, could still work for smaller grains or
grains which are more fluffy.
The efficiency of CO depletion, as presented in \citet{kri18}, should be investigated for the vertical shear instability in the
future.
\subsection{Influence on pebble accretion}
The growth of planetesimals and planetary embryos by the accretion of pebbles is an important part
of planet formation \citep{joh15}. For efficient pebble accretion the pebbles'
scale height should be smaller than the Hill sphere of the planet, $H_d <
R_H$. The aspect ratio of our millimeter particles $h_d=0.034$ means that a
planet at $r_p=50$~au has Hill radius $R_H = r_p \left ( q_p \over 3 \right )^{1/3}$ exceeding the particles' scale height if the planet's mass
  is greater than $q_p=1.2\times 10^{-4}$~times the stellar mass.  For our
  0.5-$M_\odot$ star this translates to efficient accretion of
  millimeter-sized pebbles by planets with more than $20M_\oplus$. 
\citet{pic18} investigated the pebble accretion efficiency in VSI turbulent disks
and found a peak for Stokes number unity grains because of their more effective settling. 
Dust coagulation, fragmentation, and radial drift models predict typically a
maximum Stokes number of between $10^{-2}$ and $0.1$ \citep{tes14,ued19},
similar to our 100 $\mu$m and mm-sized grains.  For these values, pebble
accretion efficiency is reduced according to \citet{pic18}. Further study is
needed of the pebble accretion efficiency on planets embedded in VSI
turbulent disks. Our results indicate that the turbulent 
velocities induced by the VSI are likely to lead to the fragmentation of colliding mm-sized grains,
and the scale height of mm-sized grains is too large to allow for efficient
pebble accretion by terrestrial-mass planetary embryos. It is difficult to avoid concluding that the VSI is not conducive to either the
early or late phases of planet formation. Both problems, however, could be overcome if strong 
local concentrations of dust grains can arise such that the VSI is suppressed by the additional
inertia provided by the particles \citep{2019MNRAS.485.5221L}.
\subsection{What determines observed dust scale heights?}
 Good measurements of the dust scale height are so far available only
for a few of the brightest protoplanetary disks {, assuming the dust
  opacity and the relevant grain sizes are correctly modelled}.  One of these is
HL~Tau, where the dust scale height is much smaller than the expected
gas scale height, with a best fit $h_d= 0.007$ at 100~au for
millimeter-size grains \citep{pin16}.
Millimeter-sized grains' abundance is confirmed by the system's spectrum at millimeter to centimeter
wavelengths \citep{car19}.  However, to compare our
results to HL~Tau, we must allow for the differing Stokes number.  If
the grain density and size are the same, the Stokes number is
proportional to $\mathit{\rm St} \sim 1/(\rho_g h)$.
 
The gas density in the disk around HL~Tau has not been observationally
determined, so we estimate upper and lower limits.  The upper limit
comes from the Toomre $Q$ parameter, which governs when the disk is
gravitationally unstable.  As the disk shows no $m=2$ pattern or
spiral, it is likely stable against self-gravity.  Following \citet{has17},
the gas surface density at 35~au then cannot exceed 100~g~cm$^{-2}$, eight times our value.  The lower limit we derive
from HCO$^+$ observations by \citet{yen16}.  An HCO$^+$-to-H$_2$
ratio of 10$^{-8}$ \citep{cle15} would mean the HCO$^+$ column
density of 10$^{16}$~cm$^{-2}$ at this position translates to a gas
surface density of 3~g~cm$^{-2}$.  The gas scale height at 35~au from
HL~Tau is only slightly smaller than in our model, with a ratio
$(80 \mathrm{K}/30 \mathrm{K})^{1/2} \times (0.5M_\odot/1.7M_\odot)^{1/2} \approx 0.9$.  Taking all this together, for the upper gas density
limit we estimate that the Stokes number for millimeter grains at this
position in HL~Tau is 7.2 times less than in our model, which by
eq. 12 implies an even greater dust scale height of 0.072.  For the
lower gas density limit, which is around four times less than in our
model, the Stokes number would be bigger by a factor 4.5, and so the
dust scale height would be 0.017, still significantly greater than the
value of 0.007 measured for HL~Tau.
 
In both gas density limits, the strong VSI mixing would lead to a
greater dust scale height than observed in HL~Tau.  In the lower
limit, the dust-to-gas mass ratio would be near 0.1, perhaps large
enough that the dust's back-reaction on the gas could hamper VSI.
However, it is unclear how a globally-enhanced dust-to-gas ratio would
be reached in such a young system.
 
If the VSI is absent from these regions of protostellar disks, what
could be the reason?

One possibility is that magnetic activity in the disk's atmosphere could lead to
turbulence that interferes with VSI \citep{lin19,cui19}.  \citet{nel13}
found that VSI is prevented when the magnetic accretion-stress-to-pressure ratio is several times
$10^{-4}$.  Whether magnetic activity is possible at these locations is worth further investigation.

\section{Conclusion}
We have examined the implications for protoplanetary disks of the vertical
shear instability, using 3D radiation hydrodynamical simulations covering all
$360^\circ$ of azimuth at a high resolution of 70~cells per scale height.
Heating by the star's light and cooling by thermal infrared re-radiation are
included.  The star and disk have parameters typical for a T~Tauri system.
Dust particles 0.1 and 1~mm in radius are tracked to examine how the VSI
affects their motion. 

Our findings for the gas kinematics and structure are the following:
\begin{itemize}
\item The vertical shear instability is vigorous beyond 30~au from the star, but nearly absent near the domain's inner edge at 20~au, where the longer thermal relaxation timescale hinders VSI's growth.  The accretion stress rises from 20 to 30~au, leading to a dip in the gas surface density centered at 35~au, and an associated ring and gap in the distribution of the dust particles.

\item The VSI leads to flows whose saturated state yields a stress-to-pressure
  ratio $\alpha \approx 10^{-4}$.  The value of $\alpha$ rises with radius to
  a peak near 35~au, and falls only slightly below this peak in the outer
  disk.  The flows are mainly vertical, roughly axisymmetric, and cross the
  midplane with upward and downward speeds of 50 to 100~m~s$^-1$. 

\item Rossby wave instability forms a single large vortex on the inner edge of the surface density dip.  The vortex persists for the whole remainder of the run and shows no significant net migration.  Its azimuthal and radial extent are approximately 40 and 4~au.

\item Many smaller vortices are seen by the vortensity maps and through the dust concentrations they produce.  The smaller vortices are short-lived, disappearing usually within about one local orbital period.

\end{itemize}

Our findings for the dust evolution and structure based on the 100 $\mu m$
and 1mm particles are as follows:
\begin{itemize}
\item The scale height of the mm-size grains { subject to the VSI} is much larger than inferred from
  radio continuum observations of the thermal dust emission in protoplanetary
  disks. In units of the distance to the star, the scale height is 0.034,
  whereas values of 0.007 were found both in non-ideal MHD simulations \citep{flo17} and in radiative transfer modeling fit to observations of the HL Tau disk \citep{pin16}.
\item The vertical shear instability mixes particles vertically much more
  strongly than it transfers angular momentum radially.
  Fitting the grains' scale height by turbulent diffusion requires a
  coefficient 27 times greater than expected from the angular momentum
  transport rate, indicating extreme anisotropy.
  The mm-size grains' time- and space-averaged RMS vertical turbulent speed $\left<v_z^2\right>^{1/2} = 0.017 c_s$ and the eddy timescale is $0.2 \Omega^{-1}$, comparable to the velocity autocorrelation timescale of magneto-rotational turbulence \citep{tur06}.
\item The RMS turbulent speeds of the 0.1 and 1~mm grains are at least 14 and
  12~m~s$^{-1}$. Grains colliding at these speeds would fragment.
  Whether VSI drives fragmentation in these disks could in principle be
  determined by combining higher-resolution versions of the simulations with
  observations to determine the grain sizes present.
\item The pressure perturbations by the vertical shear instability reduce the
  radial drift of particles. The mm-size grains drift radially inwards at speeds of about 20~au per
  $10^4$~years at distances near 50~au which is only 16\% lower than expected in a laminar disk. While drifting, they undergo upward
  and downward motions induced by the vertical shear instability.
  Some lucky grains drift more slowly, and some halt their drift when they become trapped inside the large vortex.
\end{itemize}
The large dust scale height produced by VSI turbulence appears incompatible
with the small scale height measured in the HL Tau disk.
The VSI's efficient vertical mixing would lead to the inclined disk's rings being projected one on top of the next, making the surface brightness profile
much smoother than observed \citep{pin16}.
Further work is needed to identify the dominant angular transport mechanism,
a crucial ingredient for our understanding of gas and dust evolution and planet formation in these disks.

\section*{Acknowledgments}
We thank Chris Ormel for helpful discussions on the relative velocities between
grains. Parallel computations have been performed on the pleiades supercomputer at
NASA. This research was carried out in part at the Jet Propulsion Laboratory,
California Institute of Technology, under a contract with the National
Aeronautics and Space Administration and with the support of the NASA
Exoplanet Research program via grant 14\-XRP14\_2\-0153. M. F. has received
funding from the European Research Council (ERC) under the European Unions Horizon 2020 research
and innovation programme (grant agreement No. 757957). N.J.T. was supported by
the NASA Exoplanets Research Program through grant 17\-XRP17\_2\-0081. Richard Nelson acknowledges support from STFC through the grants ST/P000592/1 and ST/M001202/1. This research was supported in part by the National Science Foundation under Grant No. NSF PHY-1125915. Copyright 2017 California Institute of Technology. Government sponsorship acknowledged. 
%\appendix
% 
%
\bibliographystyle{aasjournal}
\bibliography{VSI}

\begin{thebibliography}{}
\expandafter\ifx\csname natexlab\endcsname\relax\def\natexlab#1{#1}\fi
\providecommand{\url}[1]{\href{#1}{#1}}
\providecommand{\dodoi}[1]{doi:~\href{http://doi.org/#1}{\nolinkurl{#1}}}
\providecommand{\doeprint}[1]{\href{http://ascl.net/#1}{\nolinkurl{http://ascl.net/#1}}}
\providecommand{\doarXiv}[1]{\href{https://arxiv.org/abs/#1}{\nolinkurl{https://arxiv.org/abs/#1}}}

\bibitem[{{ALMA Partnership} {et~al.}(2015){ALMA Partnership}, {Brogan},
  {P{\'e}rez}, {Hunter}, {Dent}, {Hales}, {Hills}, {Corder}, {Fomalont},
  {Vlahakis}, {Asaki}, {Barkats}, {Hirota}, {Hodge}, {Impellizzeri}, {Kneissl},
  {Liuzzo}, {Lucas}, {Marcelino}, {Matsushita}, {Nakanishi}, {Phillips},
  {Richards}, {Toledo}, {Aladro}, {Broguiere}, {Cortes}, {Cortes}, {Espada},
  {Galarza}, {Garcia-Appadoo}, {Guzman-Ramirez}, {Humphreys}, {Jung}, {Kameno},
  {Laing}, {Leon}, {Marconi}, {Mignano}, {Nikolic}, {Nyman}, {Radiszcz},
  {Remijan}, {Rod{\'o}n}, {Sawada}, {Takahashi}, {Tilanus}, {Vila Vilaro},
  {Watson}, {Wiklind}, {Akiyama}, {Chapillon}, {de Gregorio-Monsalvo}, {Di
  Francesco}, {Gueth}, {Kawamura}, {Lee}, {Nguyen Luong}, {Mangum}, {Pietu},
  {Sanhueza}, {Saigo}, {Takakuwa}, {Ubach}, {van Kempen}, {Wootten},
  {Castro-Carrizo}, {Francke}, {Gallardo}, {Garcia}, {Gonzalez}, {Hill},
  {Kaminski}, {Kurono}, {Liu}, {Lopez}, {Morales}, {Plarre}, {Schieven},
  {Testi}, {Videla}, {Villard}, {Andreani}, {Hibbard}, \& {Tatematsu}}]{par15}
{ALMA Partnership}, {Brogan}, C.~L., {P{\'e}rez}, L.~M., {et~al.} 2015, \apjl,
  808, L3, \dodoi{10.1088/2041-8205/808/1/L3}

\bibitem[{Armitage(2019)}]{arm19}
Armitage, P.~J. 2019, Physical Processes in Protoplanetary Disks, ed.
  M.~Audard, M.~R. Meyer, \& Y.~Alibert (Berlin, Heidelberg: Springer Berlin
  Heidelberg), 1--150, \dodoi{10.1007/978-3-662-58687-7_1}

\bibitem[{{Bae} {et~al.}(2016){Bae}, {Nelson}, {Hartmann}, \&
  {Richard}}]{bae16}
{Bae}, J., {Nelson}, R.~P., {Hartmann}, L., \& {Richard}, S. 2016, \apj, 829,
  13, \dodoi{10.3847/0004-637X/829/1/13}

\bibitem[{{Barge} {et~al.}(2016){Barge}, {Richard}, \& {Le Diz{\`e}s}}]{bar16}
{Barge}, P., {Richard}, S., \& {Le Diz{\`e}s}, S. 2016, \aap, 592, A136,
  \dodoi{10.1051/0004-6361/201628381}

\bibitem[{{Cameron} \& {Pine}(1973)}]{cam73}
{Cameron}, A.~G.~W., \& {Pine}, M.~R. 1973, \icarus, 18, 377,
  \dodoi{10.1016/0019-1035(73)90152-8}

\bibitem[{{Carrasco-Gonz{\'a}lez} {et~al.}(2019){Carrasco-Gonz{\'a}lez},
  {Sierra}, {Flock}, {Zhu}, {Henning}, {Chandler}, {Galv{\'a}n-Madrid},
  {Mac{\'\i}as}, {Anglada}, {Linz}, {Osorio}, {Rodr{\'\i}guez}, {Testi},
  {Torrelles}, {P{\'e}rez}, \& {Liu}}]{car19}
{Carrasco-Gonz{\'a}lez}, C., {Sierra}, A., {Flock}, M., {et~al.} 2019, \apj,
  883, 71, \dodoi{10.3847/1538-4357/ab3d33}

\bibitem[{{Chen} \& {Lin}(2020)}]{che20}
{Chen}, K., \& {Lin}, M.-K. 2020, arXiv e-prints, arXiv:2002.07188.
\newblock \doarXiv{2002.07188}

\bibitem[{{Chiang} \& {Goldreich}(1997)}]{chi97}
{Chiang}, E.~I., \& {Goldreich}, P. 1997, \apj, 490, 368,
  \dodoi{10.1086/304869}

\bibitem[{{Cleeves} {et~al.}(2015){Cleeves}, {Bergin}, {Qi}, {Adams}, \&
  {{\"O}berg}}]{cle15}
{Cleeves}, L.~I., {Bergin}, E.~A., {Qi}, C., {Adams}, F.~C., \& {{\"O}berg},
  K.~I. 2015, \apj, 799, 204, \dodoi{10.1088/0004-637X/799/2/204}

\bibitem[{{Cui} \& {Bai}(2019)}]{cui19}
{Cui}, C., \& {Bai}, X.-N. 2019, arXiv e-prints, arXiv:1912.02941.
\newblock \doarXiv{1912.02941}

\bibitem[{{Decampli} {et~al.}(1978){Decampli}, {Cameron}, {Bodenheimer}, \&
  {Black}}]{dec78}
{Decampli}, W.~M., {Cameron}, A.~G.~W., {Bodenheimer}, P., \& {Black}, D.~C.
  1978, \apj, 223, 854, \dodoi{10.1086/156318}

\bibitem[{{Dubrulle} {et~al.}(1995){Dubrulle}, {Morfill}, \& {Sterzik}}]{dub95}
{Dubrulle}, B., {Morfill}, G., \& {Sterzik}, M. 1995, \icarus, 114, 237,
  \dodoi{10.1006/icar.1995.1058}

\bibitem[{{Dullemond} \& {Dominik}(2004)}]{dul04}
{Dullemond}, C.~P., \& {Dominik}, C. 2004, \aap, 417, 159,
  \dodoi{10.1051/0004-6361:20031768}

\bibitem[{{Dullemond} \& {Penzlin}(2018)}]{dul18a}
{Dullemond}, C.~P., \& {Penzlin}, A.~B.~T. 2018, \aap, 609, A50,
  \dodoi{10.1051/0004-6361/201731878}

\bibitem[{{Flock} {et~al.}(2013){Flock}, {Fromang}, {Gonz{\'a}lez}, \&
  {Commer{\c c}on}}]{flo13}
{Flock}, M., {Fromang}, S., {Gonz{\'a}lez}, M., \& {Commer{\c c}on}, B. 2013,
  \aap, 560, A43, \dodoi{10.1051/0004-6361/201322451}

\bibitem[{{Flock} {et~al.}(2016){Flock}, {Fromang}, {Turner}, \&
  {Benisty}}]{flo16}
{Flock}, M., {Fromang}, S., {Turner}, N.~J., \& {Benisty}, M. 2016, \apj, 827,
  144, \dodoi{10.3847/0004-637X/827/2/144}

\bibitem[{{Flock} {et~al.}(2017{\natexlab{a}}){Flock}, {Fromang}, {Turner}, \&
  {Benisty}}]{flo17a}
---. 2017{\natexlab{a}}, \apj, 835, 230, \dodoi{10.3847/1538-4357/835/2/230}

\bibitem[{{Flock} {et~al.}(2017{\natexlab{b}}){Flock}, {Nelson}, {Turner},
  {Bertrang}, {Carrasco-Gonz{\'a}lez}, {Henning}, {Lyra}, \& {Teague}}]{flo17}
{Flock}, M., {Nelson}, R.~P., {Turner}, N.~J., {et~al.} 2017{\natexlab{b}},
  \apj, 850, 131, \dodoi{10.3847/1538-4357/aa943f}

\bibitem[{{Flock} {et~al.}(2015){Flock}, {Ruge}, {Dzyurkevich}, {Henning},
  {Klahr}, \& {Wolf}}]{flo15}
{Flock}, M., {Ruge}, J.~P., {Dzyurkevich}, N., {et~al.} 2015, \aap, 574, A68,
  \dodoi{10.1051/0004-6361/201424693}

\bibitem[{{Fricke}(1968)}]{fri68}
{Fricke}, K. 1968, \zap, 68, 317

\bibitem[{{Fromang} \& {Lesur}(2019)}]{fro19}
{Fromang}, S., \& {Lesur}, G. 2019, in EAS Publications Series, Vol.~82, EAS
  Publications Series, 391--413, \dodoi{10.1051/eas/1982035}

\bibitem[{{Fromang} \& {Papaloizou}(2006)}]{fro06b}
{Fromang}, S., \& {Papaloizou}, J. 2006, \aap, 452, 751,
  \dodoi{10.1051/0004-6361:20054612}

\bibitem[{{Goldreich} \& {Schubert}(1967)}]{gol67}
{Goldreich}, P., \& {Schubert}, G. 1967, \apj, 150, 571, \dodoi{10.1086/149360}

\bibitem[{{Gr{\"a}fe} {et~al.}(2013){Gr{\"a}fe}, {Wolf}, {Guilloteau},
  {Dutrey}, {Stapelfeldt}, {Pontoppidan}, \& {Sauter}}]{gra13}
{Gr{\"a}fe}, C., {Wolf}, S., {Guilloteau}, S., {et~al.} 2013, \aap, 553, A69,
  \dodoi{10.1051/0004-6361/201220720}

\bibitem[{{Hartmann} \& {Bae}(2018)}]{har18}
{Hartmann}, L., \& {Bae}, J. 2018, \mnras, 474, 88,
  \dodoi{10.1093/mnras/stx2775}

\bibitem[{{Hasegawa} {et~al.}(2017){Hasegawa}, {Okuzumi}, {Flock}, \&
  {Turner}}]{has17}
{Hasegawa}, Y., {Okuzumi}, S., {Flock}, M., \& {Turner}, N.~J. 2017, \apj, 845,
  31, \dodoi{10.3847/1538-4357/aa7d55}

\bibitem[{{Johansen} \& {Klahr}(2005)}]{joh05}
{Johansen}, A., \& {Klahr}, H. 2005, \apj, 634, 1353, \dodoi{10.1086/497118}

\bibitem[{{Johansen} {et~al.}(2015){Johansen}, {Mac Low}, {Lacerda}, \&
  {Bizzarro}}]{joh15}
{Johansen}, A., {Mac Low}, M.-M., {Lacerda}, P., \& {Bizzarro}, M. 2015,
  Science Advances, 1, 1500109, \dodoi{10.1126/sciadv.1500109}

\bibitem[{{Klahr} \& {Hubbard}(2014)}]{kla14}
{Klahr}, H., \& {Hubbard}, A. 2014, \apj, 788, 21,
  \dodoi{10.1088/0004-637X/788/1/21}

\bibitem[{{Klahr} \& {Bodenheimer}(2003)}]{kla03}
{Klahr}, H.~H., \& {Bodenheimer}, P. 2003, \apj, 582, 869,
  \dodoi{10.1086/344743}

\bibitem[{{Krijt} {et~al.}(2018){Krijt}, {Schwarz}, {Bergin}, \&
  {Ciesla}}]{kri18}
{Krijt}, S., {Schwarz}, K.~R., {Bergin}, E.~A., \& {Ciesla}, F.~J. 2018, \apj,
  864, 78, \dodoi{10.3847/1538-4357/aad69b}

\bibitem[{{Latter} \& {Papaloizou}(2018)}]{lat18}
{Latter}, H.~N., \& {Papaloizou}, J. 2018, \mnras, 474, 3110,
  \dodoi{10.1093/mnras/stx3031}

\bibitem[{{Lesur} \& {Papaloizou}(2009)}]{les09}
{Lesur}, G., \& {Papaloizou}, J.~C.~B. 2009, \aap, 498, 1,
  \dodoi{10.1051/0004-6361/200811577}

\bibitem[{{Lesur} \& {Papaloizou}(2010)}]{les10}
---. 2010, \aap, 513, A60, \dodoi{10.1051/0004-6361/200913594}

\bibitem[{{Lesur} \& {Latter}(2016)}]{les16}
{Lesur}, G.~R.~J., \& {Latter}, H. 2016, \mnras, 462, 4549,
  \dodoi{10.1093/mnras/stw2172}

\bibitem[{{Levermore} \& {Pomraning}(1981)}]{lev81}
{Levermore}, C.~D., \& {Pomraning}, G.~C. 1981, \apj, 248, 321,
  \dodoi{10.1086/159157}

\bibitem[{{Li} {et~al.}(2000){Li}, {Finn}, {Lovelace}, \& {Colgate}}]{li00}
{Li}, H., {Finn}, J.~M., {Lovelace}, R.~V.~E., \& {Colgate}, S.~A. 2000, \apj,
  533, 1023, \dodoi{10.1086/308693}

\bibitem[{{Lin}(2019{\natexlab{a}})}]{lin19}
{Lin}, M.-K. 2019{\natexlab{a}}, \mnras, 485, 5221,
  \dodoi{10.1093/mnras/stz701}

\bibitem[{{Lin}(2019{\natexlab{b}})}]{2019MNRAS.485.5221L}
---. 2019{\natexlab{b}}, \mnras, 485, 5221, \dodoi{10.1093/mnras/stz701}

\bibitem[{{Lin} \& {Youdin}(2015)}]{lin15}
{Lin}, M.-K., \& {Youdin}, A.~N. 2015, \apj, 811, 17,
  \dodoi{10.1088/0004-637X/811/1/17}

\bibitem[{{Liu} {et~al.}(2012){Liu}, {Madlener}, {Wolf}, {Wang}, \&
  {Ruge}}]{liu12}
{Liu}, Y., {Madlener}, D., {Wolf}, S., {Wang}, H., \& {Ruge}, J.~P. 2012, \aap,
  546, A7, \dodoi{10.1051/0004-6361/201219336}

\bibitem[{{Liu} {et~al.}(2017){Liu}, {Henning}, {Carrasco-Gonz{\'a}lez},
  {Chandler}, {Linz}, {Birnstiel}, {van Boekel}, {P{\'e}rez}, {Flock}, {Testi},
  {Rodr{\'\i}guez}, \& {Galv{\'a}n-Madrid}}]{liu17}
{Liu}, Y., {Henning}, T., {Carrasco-Gonz{\'a}lez}, C., {et~al.} 2017, \aap,
  607, A74, \dodoi{10.1051/0004-6361/201629786}

\bibitem[{{Lyra}(2013)}]{lyr13}
{Lyra}, W. 2013, in European Physical Journal Web of Conferences, Vol.~46,
  European Physical Journal Web of Conferences, 04003,
  \dodoi{10.1051/epjconf/20134604003}

\bibitem[{{Lyra}(2014)}]{lyr14}
{Lyra}, W. 2014, \apj, 789, 77, \dodoi{10.1088/0004-637X/789/1/77}

\bibitem[{{Lyra} {et~al.}(2009){Lyra}, {Johansen}, {Zsom}, {Klahr}, \&
  {Piskunov}}]{lyr09}
{Lyra}, W., {Johansen}, A., {Zsom}, A., {Klahr}, H., \& {Piskunov}, N. 2009,
  \aap, 497, 869, \dodoi{10.1051/0004-6361/200811265}

\bibitem[{{Lyra} \& {Umurhan}(2019)}]{lyr19}
{Lyra}, W., \& {Umurhan}, O.~M. 2019, \pasp, 131, 072001,
  \dodoi{10.1088/1538-3873/aaf5ff}

\bibitem[{{Madlener} {et~al.}(2012){Madlener}, {Wolf}, {Dutrey}, \&
  {Guilloteau}}]{mad12}
{Madlener}, D., {Wolf}, S., {Dutrey}, A., \& {Guilloteau}, S. 2012, \aap, 543,
  A81, \dodoi{10.1051/0004-6361/201117615}

\bibitem[{{Malygin} {et~al.}(2017){Malygin}, {Klahr}, {Semenov}, {Henning}, \&
  {Dullemond}}]{mal17}
{Malygin}, M.~G., {Klahr}, H., {Semenov}, D., {Henning}, T., \& {Dullemond},
  C.~P. 2017, \aap, 605, A30, \dodoi{10.1051/0004-6361/201629933}

\bibitem[{{Manger} \& {Klahr}(2018)}]{man18}
{Manger}, N., \& {Klahr}, H. 2018, \mnras, 480, 2125,
  \dodoi{10.1093/mnras/sty1909}

\bibitem[{{Marcus} {et~al.}(2015){Marcus}, {Pei}, {Jiang}, {Barranco},
  {Hassanzadeh}, \& {Lecoanet}}]{mar15}
{Marcus}, P.~S., {Pei}, S., {Jiang}, C.-H., {et~al.} 2015, \apj, 808, 87,
  \dodoi{10.1088/0004-637X/808/1/87}

\bibitem[{{Marcus} {et~al.}(2013){Marcus}, {Pei}, {Jiang}, \&
  {Hassanzadeh}}]{mar13}
{Marcus}, P.~S., {Pei}, S., {Jiang}, C.-H., \& {Hassanzadeh}, P. 2013, Physical
  Review Letters, 111, 084501, \dodoi{10.1103/PhysRevLett.111.084501}

\bibitem[{{Masset}(2000)}]{mas00}
{Masset}, F. 2000, \aaps, 141, 165, \dodoi{10.1051/aas:2000116}

\bibitem[{{Meheut} {et~al.}(2013){Meheut}, {Lovelace}, \& {Lai}}]{meh13}
{Meheut}, H., {Lovelace}, R.~V.~E., \& {Lai}, D. 2013, \mnras, 430, 1988,
  \dodoi{10.1093/mnras/stt022}

\bibitem[{{Mignone} {et~al.}(2007){Mignone}, {Bodo}, {Massaglia}, {Matsakos},
  {Tesileanu}, {Zanni}, \& {Ferrari}}]{mig07}
{Mignone}, A., {Bodo}, G., {Massaglia}, S., {et~al.} 2007, \apjs, 170, 228,
  \dodoi{10.1086/513316}

\bibitem[{{Mignone} {et~al.}(2012){Mignone}, {Zanni}, {Tzeferacos}, {van
  Straalen}, {Colella}, \& {Bodo}}]{mig12a}
{Mignone}, A., {Zanni}, C., {Tzeferacos}, P., {et~al.} 2012, \apjs, 198, 7,
  \dodoi{10.1088/0067-0049/198/1/7}

\bibitem[{{Nelson} {et~al.}(2013){Nelson}, {Gressel}, \& {Umurhan}}]{nel13}
{Nelson}, R.~P., {Gressel}, O., \& {Umurhan}, O.~M. 2013, \mnras, 435, 2610,
  \dodoi{10.1093/mnras/stt1475}

\bibitem[{Paardekooper {et~al.}(2010)Paardekooper, Lesur, \&
  Papaloizou}]{paa10}
Paardekooper, S.-J., Lesur, G., \& Papaloizou, J. C.~B. 2010, The Astrophysical
  Journal, 725, 146, \dodoi{10.1088/0004-637x/725/1/146}

\bibitem[{{Papaloizou} \& {Pringle}(1984)}]{pap84}
{Papaloizou}, J.~C.~B., \& {Pringle}, J.~E. 1984, \mnras, 208, 721,
  \dodoi{10.1093/mnras/208.4.721}

\bibitem[{{Pfeil} \& {Klahr}(2019)}]{pfe19}
{Pfeil}, T., \& {Klahr}, H. 2019, \apj, 871, 150,
  \dodoi{10.3847/1538-4357/aaf962}

\bibitem[{{Picogna} {et~al.}(2018){Picogna}, {Stoll}, \& {Kley}}]{pic18}
{Picogna}, G., {Stoll}, M. H.~R., \& {Kley}, W. 2018, \aap, 616, A116,
  \dodoi{10.1051/0004-6361/201732523}

\bibitem[{{Pinte} {et~al.}(2016){Pinte}, {Dent}, {M{\'e}nard}, {Hales}, {Hill},
  {Cortes}, \& {de Gregorio-Monsalvo}}]{pin16}
{Pinte}, C., {Dent}, W.~R.~F., {M{\'e}nard}, F., {et~al.} 2016, \apj, 816, 25,
  \dodoi{10.3847/0004-637X/816/1/25}

\bibitem[{{Richard} {et~al.}(2016){Richard}, {Nelson}, \& {Umurhan}}]{ric16}
{Richard}, S., {Nelson}, R.~P., \& {Umurhan}, O.~M. 2016, \mnras, 456, 3571,
  \dodoi{10.1093/mnras/stv2898}

\bibitem[{{Sauter} {et~al.}(2009){Sauter}, {Wolf}, {Launhardt}, {Padgett},
  {Stapelfeldt}, {Pinte}, {Duch{\^e}ne}, {M{\'e}nard}, {McCabe}, {Pontoppidan},
  {Dunham}, {Bourke}, \& {Chen}}]{sau09}
{Sauter}, J., {Wolf}, S., {Launhardt}, R., {et~al.} 2009, \aap, 505, 1167,
  \dodoi{10.1051/0004-6361/200912397}

\bibitem[{{Sch{\"a}fer} {et~al.}(2020){Sch{\"a}fer}, {Johansen}, \&
  {Banerjee}}]{sch20}
{Sch{\"a}fer}, U., {Johansen}, A., \& {Banerjee}, R. 2020, arXiv e-prints,
  arXiv:2002.07185.
\newblock \doarXiv{2002.07185}

\bibitem[{{Schegerer} {et~al.}(2009){Schegerer}, {Wolf}, {Hummel}, {Quanz}, \&
  {Richichi}}]{sch09}
{Schegerer}, A.~A., {Wolf}, S., {Hummel}, C.~A., {Quanz}, S.~P., \& {Richichi},
  A. 2009, \aap, 502, 367, \dodoi{10.1051/0004-6361/200810782}

\bibitem[{{Schegerer} {et~al.}(2008){Schegerer}, {Wolf}, {Ratzka}, \&
  {Leinert}}]{sch08}
{Schegerer}, A.~A., {Wolf}, S., {Ratzka}, T., \& {Leinert}, C. 2008, \aap, 478,
  779, \dodoi{10.1051/0004-6361:20077049}

\bibitem[{{Schr{\"a}pler} \& {Henning}(2004)}]{sch04}
{Schr{\"a}pler}, R., \& {Henning}, T. 2004, \apj, 614, 960,
  \dodoi{10.1086/423831}

\bibitem[{{Seligman} \& {Laughlin}(2017)}]{sel17}
{Seligman}, D., \& {Laughlin}, G. 2017, \apj, 848, 54,
  \dodoi{10.3847/1538-4357/aa8e45}

\bibitem[{{Stoll} \& {Kley}(2014)}]{sto14}
{Stoll}, M.~H.~R., \& {Kley}, W. 2014, \aap, 572, A77,
  \dodoi{10.1051/0004-6361/201424114}

\bibitem[{{Stoll} \& {Kley}(2016)}]{sto16}
---. 2016, \aap, 594, A57, \dodoi{10.1051/0004-6361/201527716}

\bibitem[{{Stoll} {et~al.}(2017){Stoll}, {Kley}, \& {Picogna}}]{sto17}
{Stoll}, M. H.~R., {Kley}, W., \& {Picogna}, G. 2017, \aap, 599, L6,
  \dodoi{10.1051/0004-6361/201630226}

\bibitem[{{Testi} {et~al.}(2014){Testi}, {Birnstiel}, {Ricci}, {Andrews},
  {Blum}, {Carpenter}, {Dominik}, {Isella}, {Natta}, {Williams}, \&
  {Wilner}}]{tes14}
{Testi}, L., {Birnstiel}, T., {Ricci}, L., {et~al.} 2014, Protostars and
  Planets VI, 339, \dodoi{10.2458/azu_uapress_9780816531240-ch015}

\bibitem[{{Turner} {et~al.}(2014){Turner}, {Benisty}, {Dullemond}, \&
  {Hirose}}]{tur14}
{Turner}, N.~J., {Benisty}, M., {Dullemond}, C.~P., \& {Hirose}, S. 2014, \apj,
  780, 42, \dodoi{10.1088/0004-637X/780/1/42}

\bibitem[{{Turner} {et~al.}(2006){Turner}, {Willacy}, {Bryden}, \&
  {Yorke}}]{tur06}
{Turner}, N.~J., {Willacy}, K., {Bryden}, G., \& {Yorke}, H.~W. 2006, \apj,
  639, 1218, \dodoi{10.1086/499486}

\bibitem[{{Ueda} {et~al.}(2019){Ueda}, {Flock}, \& {Okuzumi}}]{ued19}
{Ueda}, T., {Flock}, M., \& {Okuzumi}, S. 2019, \apj, 871, 10,
  \dodoi{10.3847/1538-4357/aaf3a1}

\bibitem[{{Umurhan} {et~al.}(2019){Umurhan}, {Estrada}, \& {Cuzzi}}]{umu19}
{Umurhan}, O.~M., {Estrada}, P.~R., \& {Cuzzi}, J.~N. 2019, arXiv e-prints,
  arXiv:1906.05371.
\newblock \doarXiv{1906.05371}

\bibitem[{{Umurhan} {et~al.}(2016){Umurhan}, {Shariff}, \& {Cuzzi}}]{umu16b}
{Umurhan}, O.~M., {Shariff}, K., \& {Cuzzi}, J.~N. 2016, \apj, 830, 95,
  \dodoi{10.3847/0004-637X/830/2/95}

\bibitem[{{Urpin} \& {Brandenburg}(1998)}]{urp98}
{Urpin}, V., \& {Brandenburg}, A. 1998, \mnras, 294, 399,
  \dodoi{10.1046/j.1365-8711.1998.01118.x}

\bibitem[{{van Boekel} {et~al.}(2017){van Boekel}, {Henning}, {Menu}, {de
  Boer}, {Langlois}, {M{\"u}ller}, {Avenhaus}, {Boccaletti}, {Schmid},
  {Thalmann}, {Benisty}, {Dominik}, {Ginski}, {Girard}, {Gisler}, {Lobo Gomes},
  {Menard}, {Min}, {Pavlov}, {Pohl}, {Quanz}, {Rabou}, {Roelfsema}, {Sauvage},
  {Teague}, {Wildi}, \& {Zurlo}}]{vanb17}
{van Boekel}, R., {Henning}, T., {Menu}, J., {et~al.} 2017, \apj, 837, 132,
  \dodoi{10.3847/1538-4357/aa5d68}

\bibitem[{{Wolf} {et~al.}(2003){Wolf}, {Padgett}, \& {Stapelfeldt}}]{wol03}
{Wolf}, S., {Padgett}, D.~L., \& {Stapelfeldt}, K.~R. 2003, \apj, 588, 373,
  \dodoi{10.1086/374041}

\bibitem[{{Wolf} {et~al.}(2008){Wolf}, {Schegerer}, {Beuther}, {Padgett}, \&
  {Stapelfeldt}}]{wol08}
{Wolf}, S., {Schegerer}, A., {Beuther}, H., {Padgett}, D.~L., \& {Stapelfeldt},
  K.~R. 2008, \apjl, 674, L101, \dodoi{10.1086/529188}

\bibitem[{{Yang} {et~al.}(2018){Yang}, {Mac Low}, \& {Johansen}}]{yan18}
{Yang}, C.-C., {Mac Low}, M.-M., \& {Johansen}, A. 2018, \apj, 868, 27,
  \dodoi{10.3847/1538-4357/aae7d4}

\bibitem[{{Yen} {et~al.}(2016){Yen}, {Liu}, {Gu}, {Hirano}, {Lee},
  {Puspitaningrum}, \& {Takakuwa}}]{yen16}
{Yen}, H.-W., {Liu}, H.~B., {Gu}, P.-G., {et~al.} 2016, \apjl, 820, L25,
  \dodoi{10.3847/2041-8205/820/2/L25}

\end{thebibliography}

\appendix

\section{Stokes number at the midplane}
\begin{figure}
  \resizebox{\hsize}{!}{\includegraphics{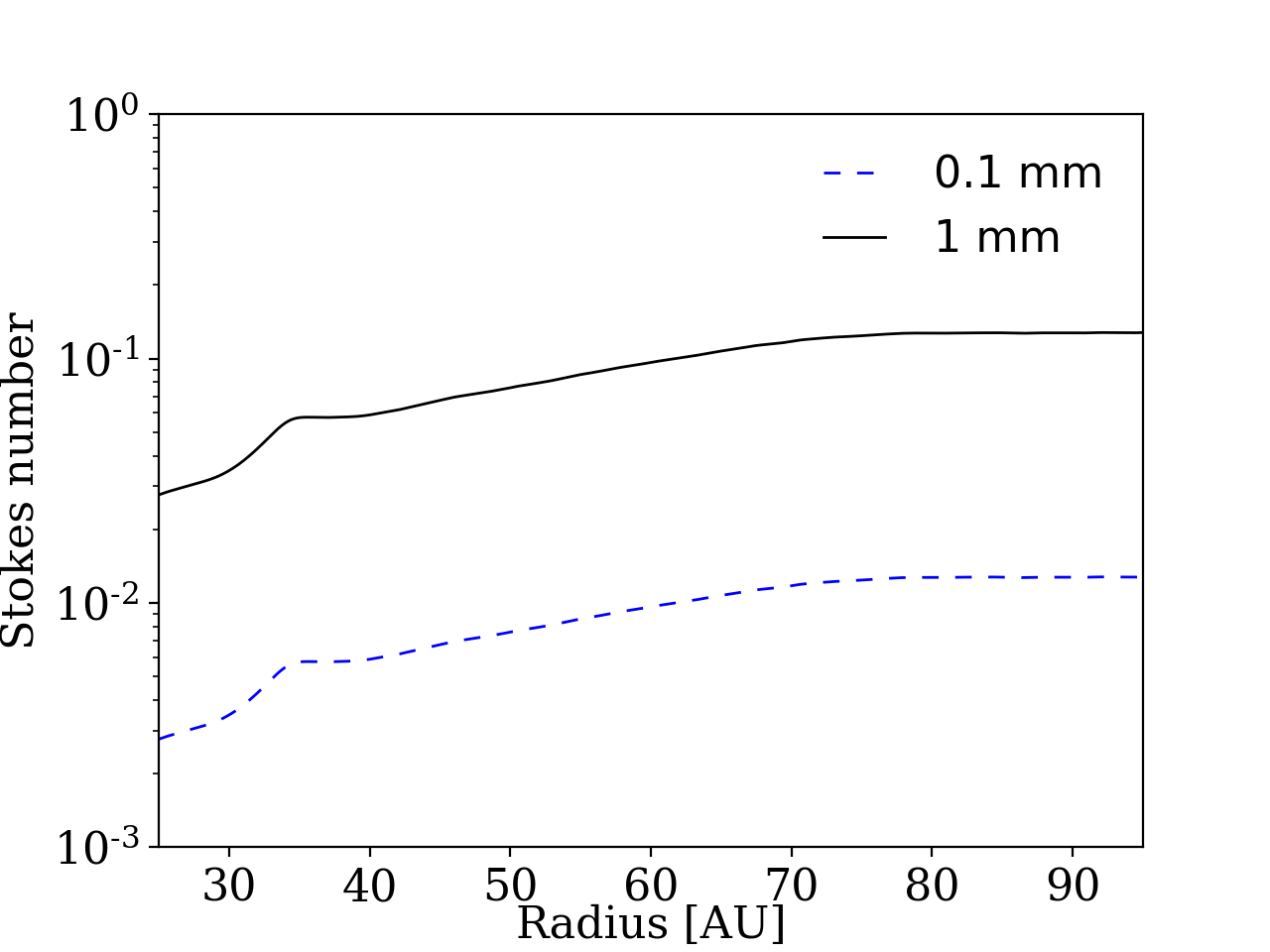}}
  \caption{Stokes number at the midplane for the 0.1 and 1mm grains after 300
    inner orbits of evolution.} 
\label{fig:stoke}
\end{figure}

The Stokes number for the 0.1 and 1 mm grains, calculated at the midplane
density, azimuthally averaged, after 300 inner orbits of gas evolution is
shown in Fig.~\ref{fig:stoke}. The 1 mm grains show midplane Stokes numbers
between $10^{-2}$ and $10^{-1}$ while the 0.1 mm grains are one order of
magnitude below. 

\section{Is the vortex caused by the Rossby-Wave-Instability?}
Following \citet{li00,lyr09,meh13}, a necessary condition for the
Rossby-Wave-Instability is an extremum in
\begin{equation}
\mathcal{L}= \frac{\mathrm{\Sigma}} { 2(\nabla \times v)_Z }  \left (\frac{ P }
{\Sigma^{-\gamma}} \right )^{2/\gamma}.
\end{equation}

\begin{figure}
  \resizebox{\hsize}{!}{\includegraphics{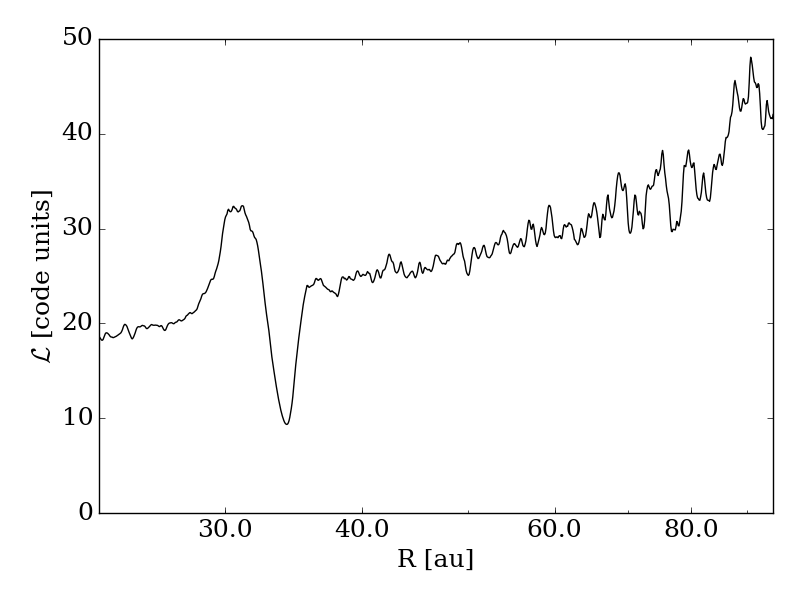}}
  \caption{Radial profile of $\mathcal{L}$, calculated at the midplane,
    azimuthally and time average for the first 100 inner orbits.} 
\label{fig:L}
\end{figure}

For our 3D model we calculate for the radial profile of $\mathcal{L}$ at the
midplane and averaging over the first 100 inner orbits. The results are shown
in Fig.~\ref{fig:L}. There is a clear peak visible at around 30 au, the
position where the vortex is formed.

\section{Evolution of short lived vortices}
\begin{figure*}
\resizebox{0.78\hsize}{!}{\includegraphics{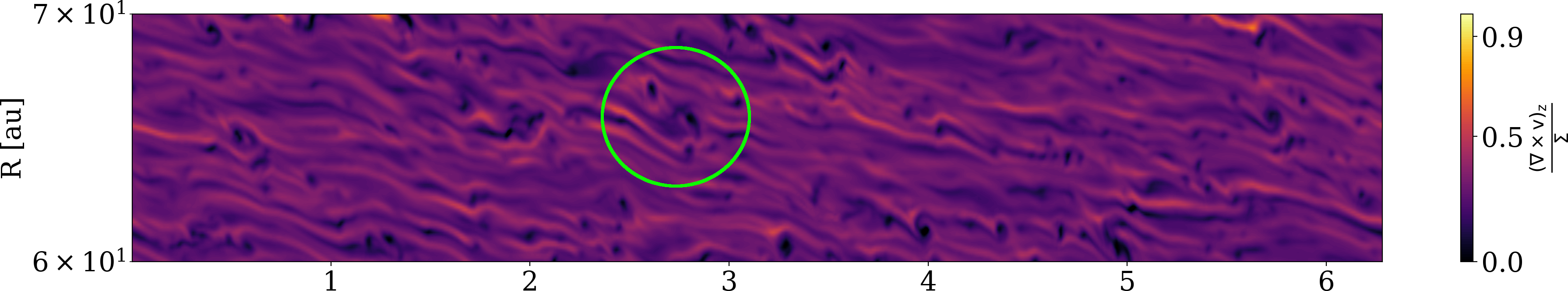}}
\resizebox{0.78\hsize}{!}{\includegraphics{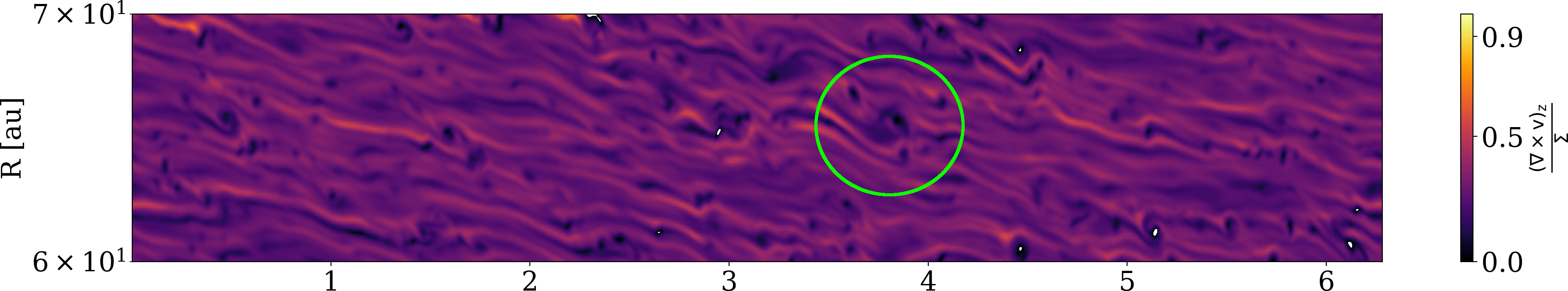}}
\resizebox{0.78\hsize}{!}{\includegraphics{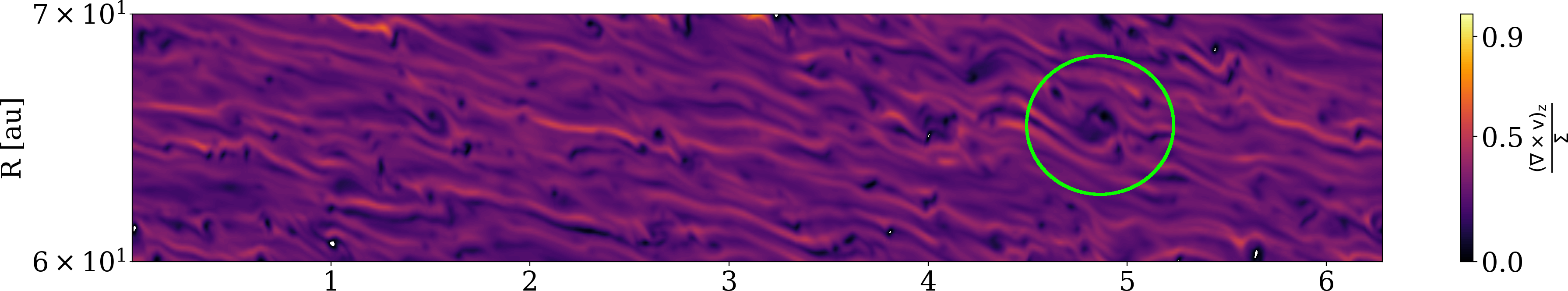}}
\resizebox{0.78\hsize}{!}{\includegraphics{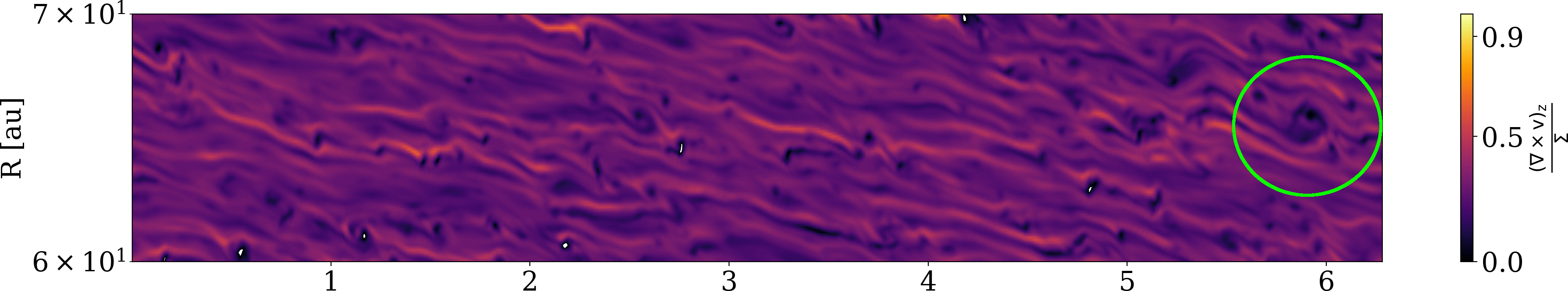}}
\resizebox{0.78\hsize}{!}{\includegraphics{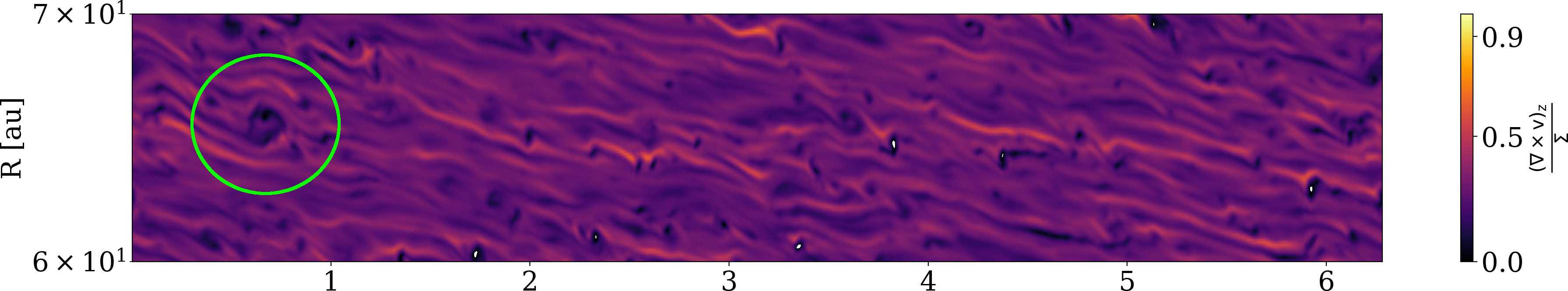}}
\resizebox{0.78\hsize}{!}{\includegraphics{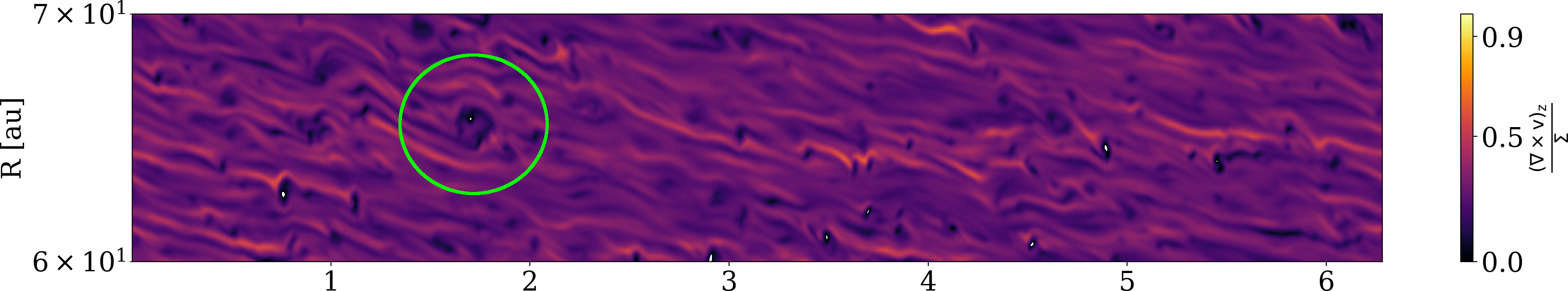}}
\resizebox{0.78\hsize}{!}{\includegraphics{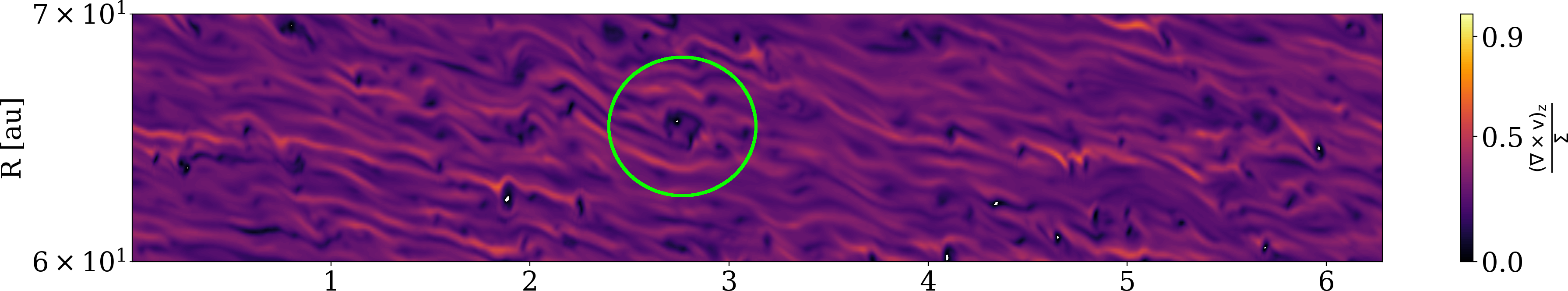}}
\resizebox{0.78\hsize}{!}{\includegraphics{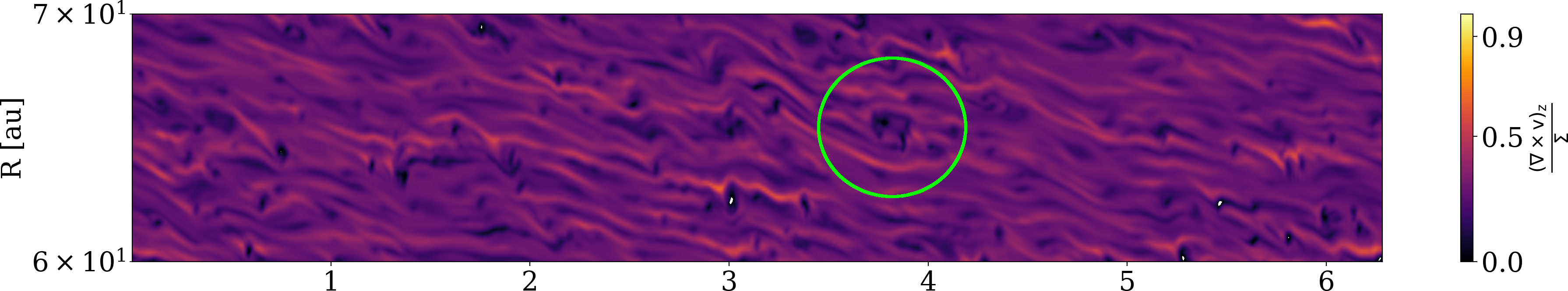}}
\resizebox{0.78\hsize}{!}{\includegraphics{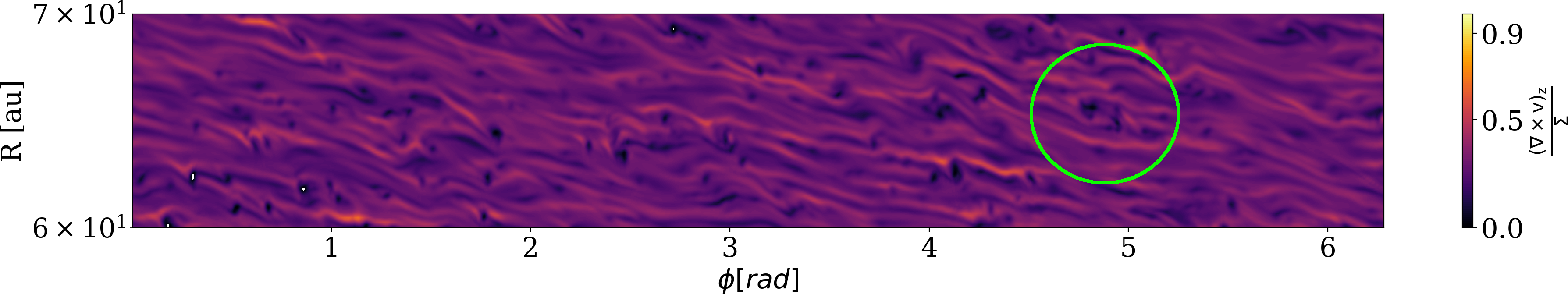}}
  \caption{Snapshot of the vortensity for a small radial region in the disk,
    taken every 0.17 local orbits at 65 au starting at 285 inner orbits. The green circle marks the position
    during formation (top) and decay (bottom).} 
\label{fig:tinyvor}
\end{figure*}

In Fig.~\ref{fig:tinyvor} we show snapshots of different times for a small
vortensity minimum which is formed and decays again after a lifetime of around 1.4 orbits. 

\section{Fully Edge-On dust comparison}
\begin{figure*}
  \resizebox{\hsize}{!}{\includegraphics{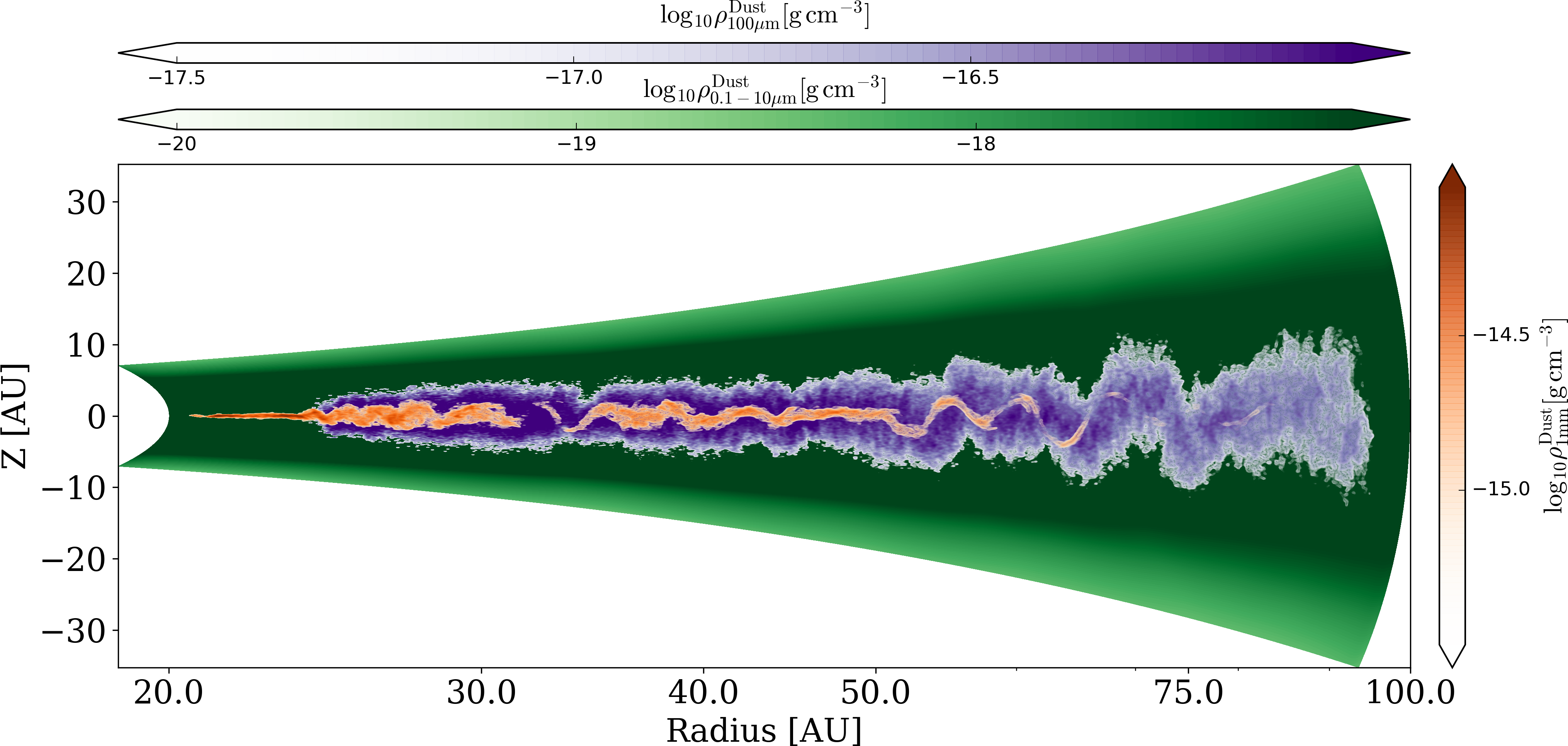}}
  \caption{{ Azimuthally averaged dust density for three representative dust
    grain size bins in the $R-Z$ plane taken after 50 inner orbits of dust
evolution.}} 
\label{fig:fullyedge}
\end{figure*}

{ To compare the dust settling for the two grain sizes we show the particle
distribution in Fig.~\ref{fig:fullyedge}.  }

\section{Relative velocities between grains}

\begin{figure}
  \resizebox{\hsize}{!}{\includegraphics{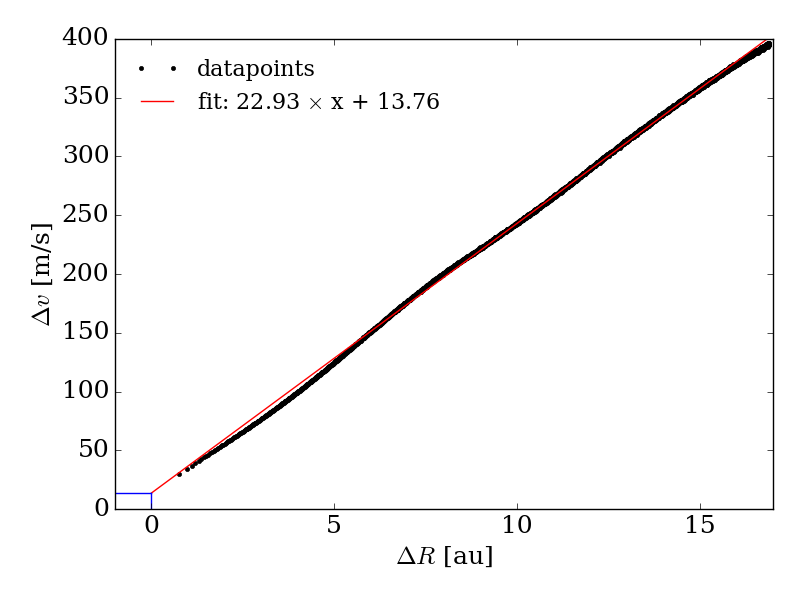}}
\caption{{Distance and relative velocity between 100 $\mu m$ grains and 1mm
    grains (black dots). A linear fit is overplotted (red line) using the grains separated by less than 17
    au. The intercept indicates that on collision we expect a relative speed of around 13.8 m/s}} 
\label{fig:dust_relvel}
\end{figure}

To further investigate the likely outcomes of collisions between grains we determine the relative velocity between the
100~$\mu m$ and 1~mm grains. For 100,000 randomly-chosen particles from each size bin we first
calculate the mutual distances and relative velocities. The resulting
$10^5 \times 10^5$ dataset is sorted over distance and then time averaged for each individual
particle. The result is shown in Fig.~\ref{fig:dust_relvel} together with a
linear fit. At collision the relative speed between the 100~$\mu m$ and
1~mm grains is around 13.8~m/s. This value matches very well the estimated turbulent speeds from section~3.6. Such collisions could
lead to fragmentation of silicate aggregate grains. However for ice covered
grains the fragmentation threshold may be
significantly higher, reaching 30 $\rm m/s$ or above \citep{ued19}. We do not
resolve the ballistic scale. At 50 au, a grain with a Stokes number of 0.05 and a velocity of 13
m/s will stop in a distance roughly 8 times smaller than our radial
grid cell size at 50~au. Resolving the stopping distance might lead to collision speeds low enough for efficient growth of ice
coated grains. It is also the case that we do resolve the characteristic
length scales of the VSI, including the length scales associated with the
coherent up and down motion of the fluid generated by the VSI linear modes,
and so the effects of these motions on the relative velocities of the grains is captured by our analysis. Future simulations 
which resolve the characteristic stopping lengths of the particles will be required to extrapolate 
the relative velocity versus distance curve with greater confidence.

\end{document}